\documentclass[%
aps,prx,reprint,%
superscriptaddress%,nofootinbib
]{revtex4-1}

\newif\ifpeter

\petertrue
\ifpeter

\usepackage{graphicx}	% Include figure files
\usepackage[bookmarks=true,%
bookmarksnumbered=true,%
bookmarkstype=toc%
]{hyperref}

\else

\usepackage[dvipdfmx]{graphicx}		% Include figure files
\usepackage[dvipdfmx,%
bookmarks=true,%
bookmarksnumbered=true,%
bookmarkstype=toc%
]{hyperref}

\fi

\usepackage[utf8]{inputenc}	% for encoding problem of biber
\usepackage{comment}
\usepackage{textcomp}
\usepackage{amssymb}
\usepackage{amsmath}  % for the \boldsymbol{...}  command
\usepackage{bm}       % for the \bm{...}  command
\usepackage{siunitx}		% for SI unit of numbers

\usepackage{hhline} % for double lines in tables
\usepackage{color}
\usepackage{colortbl}
\makeatletter
\ifpeter
\else
\fi
\makeatother

\newcommand{\current}{{I}}
\newcommand{\voltage}{{V}}

\definecolor{forestgreen}{rgb}{0.13,0.55,0.13}
\newcommand{\red}[1]{\textcolor{red}{#1}}

\definecolor{olive}{rgb}{0.0, 0.6, 0.0}

\newcommand{\ab}{\alpha_{\rm bulk}}
\newcommand{\aed}{\alpha_{\rm end}}
\newcommand{\ai}{\alpha_{\rm imp}}
\newcommand{\aA}{\alpha_{\rm A}}
\newcommand{\aB}{\alpha_{\rm B}}
\begin{document}

	\title{Strong Electron-Electron Interactions of a Tomonaga--Luttinger Liquid Observed in InAs Quantum Wires}
	\author{Yosuke \surname{Sato}}\email{sato@meso.t.u-tokyo.ac.jp}\thanks{Contributed equally to this work}
	\affiliation{Department of Applied Physics, University of Tokyo,7-3-1 Hongo, Bunkyo-ku, Tokyo 113-8656, Japan\looseness=-2}
	\author{Sadashige Matsuo}\email{matsuo@ap.t.u-tokyo.ac.jp}\thanks{Contributed equally to this work}
	\affiliation{Department of Applied Physics, University of Tokyo,7-3-1 Hongo, Bunkyo-ku, Tokyo 113-8656, Japan\looseness=-2}
	\author{Chen-Hsuan Hsu}
	\affiliation{Center for Emergent Matter Science, RIKEN, 2-1 Hirosawa, Wako-shi, Saitama 351-0198, Japan\looseness=-2}
	\author{Peter Stano}
	\affiliation{Department of Applied Physics, University of Tokyo,7-3-1 Hongo, Bunkyo-ku, Tokyo 113-8656, Japan\looseness=-2}	
	\affiliation{Center for Emergent Matter Science, RIKEN, 2-1 Hirosawa, Wako-shi, Saitama 351-0198, Japan\looseness=-2}
	\affiliation{Institute of Physics, Slovak Academy of Sciences, 845 11 Bratislava, Slovakia\looseness=-2}
	\author{Kento Ueda}
	\author{Yuusuke Takeshige}
	\affiliation{Department of Applied Physics, University of Tokyo,7-3-1 Hongo, Bunkyo-ku, Tokyo 113-8656, Japan\looseness=-2}
	\author{Hiroshi Kamata}
	\affiliation{Center for Emergent Matter Science, RIKEN, 2-1 Hirosawa, Wako-shi, Saitama 351-0198, Japan\looseness=-2}
	\author{Joon Sue Lee}
	\affiliation{California NanoSystems Institute, University of California Santa Barbara, Santa Barbara, California 93106, USA\looseness=-2}
	\author{Borzoyeh Shojaei}
	\affiliation{California NanoSystems Institute, University of California Santa Barbara, Santa Barbara, California 93106, USA\looseness=-2}
	\affiliation{Materials Engineering, University of California Santa Barbara, Santa Barbara, California 93106, USA\looseness=-2}
	\author{Kaushini Wickramasinghe}
	\affiliation{Center for Quantum Phenomena, Department of Physics,	New York University, New York, NY, 10003, USA\looseness=-2}
	\author{Javad Shabani}
	\affiliation{Center for Quantum Phenomena, Department of Physics,	New York University, New York, NY, 10003, USA\looseness=-2}
	\author{Chris Palmstr{\o}m}
	\affiliation{California NanoSystems Institute, University of California Santa Barbara, Santa Barbara, California 93106, USA\looseness=-2}
	\affiliation{Materials Engineering, University of California Santa Barbara, Santa Barbara, California 93106, USA\looseness=-2}
	\affiliation{Electrical and Computer Engineering, University of California Santa Barbara, Santa Barbara, California 93106, USA\looseness=-2}
	\author{Yasuhiro Tokura}
	\affiliation{Faculty of Pure and Applied Sciences, University of Tsukuba, Tsukuba, Ibaraki 305-8571, Japan\looseness=-2}
	\author{Daniel Loss}
	\affiliation{Center for Emergent Matter Science, RIKEN, 2-1 Hirosawa, Wako-shi, Saitama 351-0198, Japan\looseness=-2}
	\affiliation{Department of Physics, University of Basel, Klingelbergstrasse 82, CH-4056 Basel, Switzerland\looseness=-3}
	\author{Seigo Tarucha}\email{tarucha@ap.t.u-tokyo.ac.jp}
	\affiliation{Department of Applied Physics, University of Tokyo,7-3-1 Hongo, Bunkyo-ku, Tokyo 113-8656, Japan\looseness=-2}
	\affiliation{Center for Emergent Matter Science, RIKEN, 2-1 Hirosawa, Wako-shi, Saitama 351-0198, Japan\looseness=-2}

	\date{\today}
	\begin{abstract}
		We report strong electron-electron interactions in quantum wires etched from an InAs quantum well, a material known to have strong spin-orbit interactions.
		We find that the current through the wires as a function of the bias voltage and temperature follows the universal scaling behavior of a Tomonaga--Luttinger liquid.
		Using a universal scaling formula, we extract the interaction parameter and find strong electron-electron interactions, increasing as the wires become more depleted.
		We establish theoretically that spin-orbit interactions cause only minor modifications of the interaction parameter in this regime, indicating that genuinely strong electron-electron interactions are indeed achieved in the device.
		Our results suggest that etched InAs wires provide a platform with both strong electron-electron and strong spin-orbit interactions. 
	\end{abstract}

	\maketitle

\section{Introduction}

		A one-dimensional electron system displays the physics of a Tomonaga--Luttinger liquid (TLL), strikingly different to Fermi liquids in higher-dimensions.
		A spinful TLL is described by the Hamiltonian\cite{Tomonaga1950,Luttinger1963}
		\begin{align}
			H =& \sum_{\nu=\textrm{c,s}} \int \frac{\hbar dx}{2\pi} \Big\{ u_\nu g_\nu \left[\partial_x \theta_\nu (x) \right]^2 
						+\frac{u_\nu}{g_\nu} \left[ \partial_x \phi_\nu (x) \right] ^2 \Big\}.
			\label{Eq:H_TLL} 
		\end{align}
		Here, $\nu\in\{\textrm{c,s}\}$ labels the charge and spin sector, respectively, while $u_\nu$ are the velocities, and $\theta_\nu$ and $\phi_\nu$ the bosonic fields, describing the two elementary excitations\cite{Matveev2004}. 
		The electron-electron (e-e) interactions are parameterized by $g_\textrm{c}$ and $g_\textrm{s}$, numbers between 0 and 1\cite{Furusaki1993a}.
		The spin-charge separation, meaning the independence of the charge and spin sectors displayed by Eq.~\eqref{Eq:H_TLL}, appears as one of the key features of a TLL.
		
		On the other hand, the coupling of spin and charge degrees of freedom, in various forms of spin-orbit interactions (SOIs), plays an important role in semiconductors and spintronics\cite{Datta1990,Fabian2007}.
		The research on SOIs has been further accelerated 
		by predictions of the emergence of Majorana fermions in an accessible setup comprising a quantum wire with superconductivity, SOIs, and a magnetic field \cite{Lutchyn2010,Oreg2010,Klinovaja2012,Alicea2010}.
		Unfortunately, the practical realization is impeded by the incompatibility of a strong magnetic field and superconductivity.
		Recently, it has been suggested that wires with strong e-e interactions could solve this conflict by disposing of the magnetic field\cite{Klinovaja2014,Thakurathi2018}.
		More importantly, strong e-e interactions allow a realization of parafermions\cite{Klinovaja2014}, more advanced topological particles than the Majorana fermions\cite{Fendley2012, Hutter2016}.
		They rely on high-efficient Cooper pair splitting into the two quantum wires, which arises due to strong e-e interactions.
		We note that efficient Cooper pair splitting \cite{Baba2018,Ueda2018} and a transparent interface with a superconductor has been recently demonstrated in self-assembled InAs nanowires\cite{Chang2015,Krogstrup2015} and quantum well\cite{Kjaergaard2016}.
		With this outlook, providing wires with both strong e-e interactions and strong SOIs seems beneficial.		
		
		Motivated by such prospects, there appeared several theoretical works concerned with a TLL in the presence of the SOIs.
		The SOIs mix the spin and charge sectors and a rich range of phenomena was predicted, from mild modifications to a breakdown of the TLL phase\cite{Moroz2000,Moroz2000a,Hausler2001,DeMartino2001,Governale2002,Iucci2003,Governale2004,Hausler2004,Sun2007,Schulz2009,Kainaris2015a}.
		Despite active discussions in theory, there are only few experimental results about TLL physics in wires with strong SOIs. 
		Concerning InAs, we note the self-assembled nanowires\cite{Hevroni2016} and nanowire quantum point contacts\cite{Heedt2017} experiments.
		In the former, a small interaction parameter was deduced, but it remained unclear whether this was due to the SOIs, or intrinsically strong e-e interactions, or even some other physics.
		The situation contrasts to TLLs without SOIs, with a number of reports, for example on GaAs wires\cite{Tarucha1995,Asayama2002,Levy2006,Levy2012}, carbon nanotubes\cite{Bockrath1999,Postma2000,Graugnard2001,Bachtold2001,Liu2001,Lee2004}, or a more exotic realization using a quantum circuit\cite{Anthore2018}, all in which the SOIs are negligible.
		Concluding, the spin-orbit effects in TLLs are involved in theory, and remain little explored in experiments.
		
\section{Summary of the main results}
		
		Here, we investigate the TLL behavior of quantum wires fabricated in an InAs quantum well, a material known to have strong SOIs\cite{Luo1990}.
		We measure the electric current through the wires as a function of the bias voltage at various temperatures and find that the data fall onto a single curve upon rescaling.
		Such universal scaling is consistent with the TLL theory, what allows us to extract the value of the interaction parameter $g_\textrm{c}$ in Eq.~\eqref{Eq:H_TLL}.
		The extracted values reach as low as 0.16--0.28 (these minimal values are for wires close to depletion), indicating a strong-interaction regime.
		In addition to transport measurements, we provide theoretical understanding of one-dimensional systems with strong e-e interactions and SOIs.
		Overall, our results demonstrate that InAs wires offer a platform fulfilling the requirements for the realization of topological particles.
		
		\begin{figure}[t]
			\includegraphics[width=0.95\linewidth]{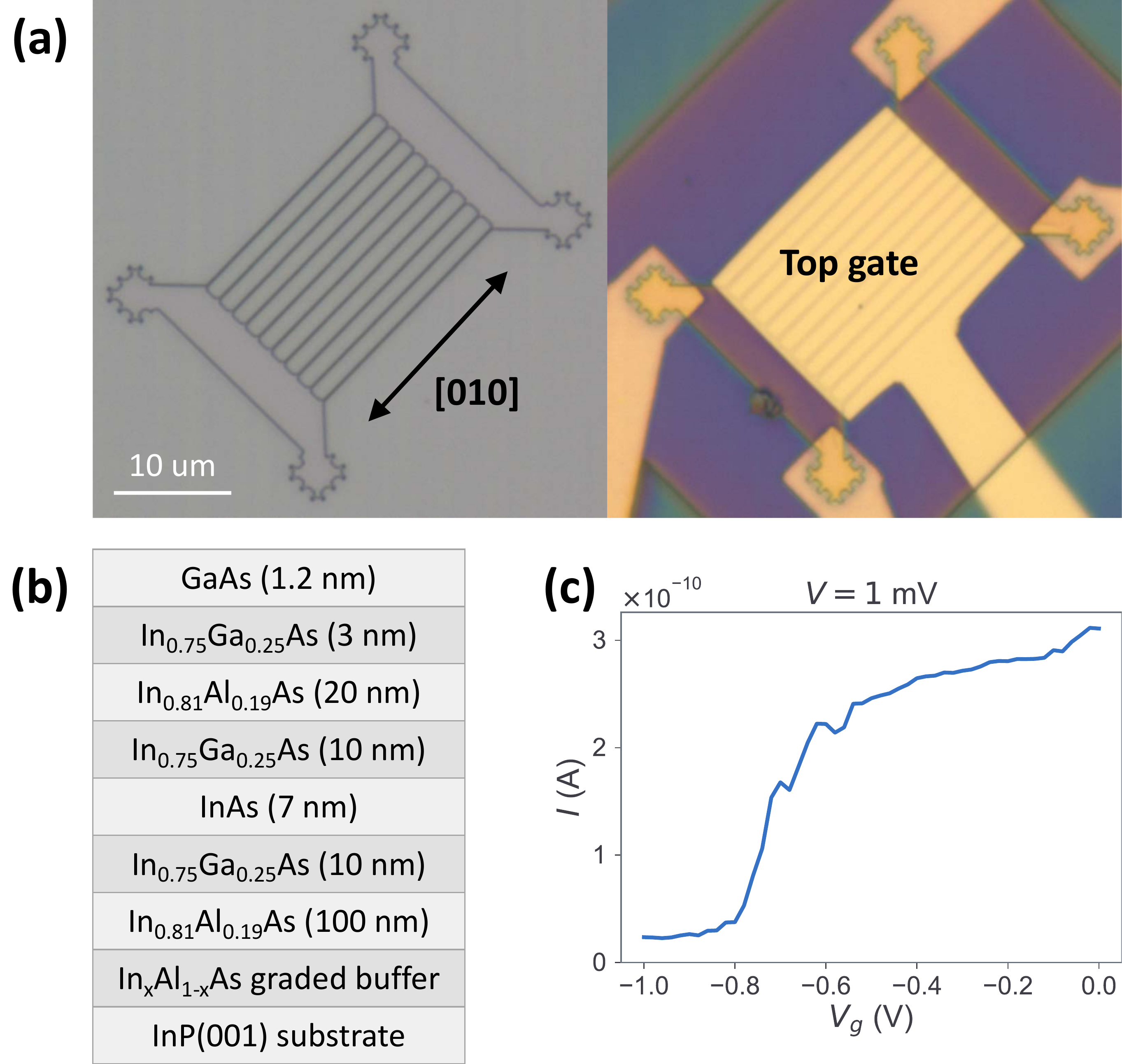}
			\caption{(a) Microscope photographs of the parallel-wire device before (left) and after (right) depositing a top gate above a cross-linked PMMA layer.
				The wires are along [010].
				(b) Schematic structure of the wafer, grown along [001].
				(c) Gate voltage dependence of the current through the wires measured at a constant source-drain voltage as given in the figure caption.
			}\label{Fig:device}
		\end{figure}
		
\section{Device}
		
		Our device, shown in Fig.~\ref{Fig:device}(a), is composed of ten parallel quantum wires, which were chemically etched from an InAs quantum well. 
		Ohmic contacts, created using Ti/Au\cite{Matsuo2017}, and a Ti/Au top-gate, deposited on top of a cross-linked PMMA serving as an insulating layer, give electrical access and control.
		A single wire has a length of \SI{20}{\micro m} and a nominal width (estimated from the microscope image) of \SI{100}{nm}.
		The stack materials of the InAs quantum well are given in Fig.~\ref{Fig:device}(b).
		Prior to measurements of the wires, the two-dimensional electron gas mobility of \SI{7.2e4}{cm/(V.s)}, electron density of \SI{3.4e11}{cm^{-2}}, and mean free path of \SI{690}{nm} were extracted from measuring a Hall-bar device at \SI{560}{mK}.
		The electric current $\current$ flowing through the parallel wires upon applying a bias voltage $\voltage$ is measured by the standard 4-terminal dc measurement. 
		These measurements are performed at temperature $T$ in the range 2--\SI{4}{K}. 
		Figure~\ref{Fig:device}(c) shows $\current$ as a function of the top gate voltage $V_g$ for a fixed $\voltage=1$ \SI{}{mV}.
		The device shows a pinch-off at about $V_g=-0.86$ V. 
		A small current remaining below that voltage is most probably due to a tunneling conductance through quantum dots formed in the disorder potential of the wires.		
		The parallel quantum wire structure reduces the total resistance such that the total current is still within the measurable range.
		Though measuring many parallel wires precludes observing conductance plateaus, it also results in averaging out the potential fluctuations from impurities and other disorder. Our guess is that such an averaging might be crucial for observing the universal scaling.

\section{Universal scaling of the current-voltage curve}
		
		Before presenting our main results, we first review the transport properties predicted by the existing theory.
		It has been established that, {\it assuming the spin-charge separation}, a current through a single TLL with several tunnel barriers (their number and positions are discussed below) displays universal scaling\cite{Bockrath1999,Balents1999}.
		Explicitly, the tunnel current is
		\begin{align}
			\current &= I_0 T^{1+\alpha} \sinh \left( \frac{\gamma eV}{2k_\textrm{B} T}\right) \left|\Gamma \left( 1+\frac{\alpha}{2}+\frac{i \gamma eV}{2\pi k_\textrm{B} T} \right) \right|^2.
			\label{Eq:USC}
		\end{align}
		Here, $I_0$ is an unspecified overall scale dependent on a typical barrier strength, $\Gamma(z)$ is the Gamma function, $e$ is the positive elementary charge, $k_\textrm{B}$ is the Boltzmann constant, and the parameters $\alpha$ and $\gamma$ depend on the number and character of the scatterers inside the wire.
		
		The expression $\gamma V$ corresponds to a voltage drop across a single scatterer. 
		Thus, in Ref.~\cite{Balents1999} which considers a single tunneling barrier, one has $\gamma = 1$.
		One can generalize this result to several, say $N$, tunneling barriers: assuming that they induce comparable resistances, a typical voltage drop over a single one will be $V/N$, from which $\gamma = 1/N$\cite{Bockrath1999,Venkataraman2006}. 
		The parameter $\gamma$ is therefore to be interpreted as the inverse number of tunnel barriers.
		
		The parameter $\alpha$ depends in an intricate way on the e-e interaction strength parameters $g_\textrm{c}$ and $g_\textrm{s}$, the SOI strength, and the number and character of the scatterers. 
		The expressions for $\alpha$ for the case of zero SOI were known before, for the case of finite one were not and we provide them here and in Ref.~\cite{Hsu2018b}.
		Extracting this parameter from the data and inferring from it the e-e interaction strength is the essence of this paper.
		Let us first describe the former task, before discussing the latter one. 
		
		The extraction of $\alpha$ is rather straightforward once the curve in Eq.~\eqref{Eq:USC} is plotted on a log-log scale. 
		Indeed, it shows different slopes for $\gamma eV$ much smaller and much larger than $k_\textrm{B} T$. For the (differential) conductance $G\equiv dI/dV$, it corresponds to a power-law $G\propto T^\alpha$ and $G \propto V^\alpha$, respectively.
		The power law in the conductance, $G \propto T^{\alpha}$, in the regime of $eV\ll k_\textrm{B} T$ was observed in numerous previous experiments\cite{Chang1996,Bockrath1999,Postma2000,Bachtold2001,Graugnard2001,Liu2001,Slot2004,Aleshin2004,Venkataraman2006,Lucot2011,Levy2012,Li2015}.
		If the universal scaling curve is obtained for a large enough range of its natural parameter, $eV/k_B T$, so that the crossover is seen, one can extract both $\gamma$ and $\alpha$. 
		This is what we do next.
		
\section{Measurement of $I-V$ curves and fit to Eq.~(2)}
		
\begin{figure}[t]
			\includegraphics[width=\linewidth]{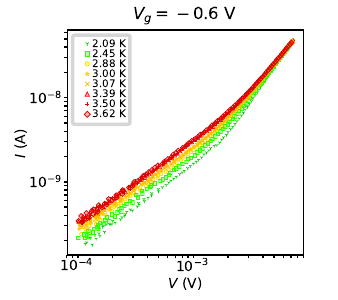}
			\caption{Current ($\current$) flowing through the wires as a function of the bias voltage ($V$) for the top gate voltage $V_g=\SI{-0.6}{V}$.} \label{Fig:I-Vcurve}
		\end{figure}

\begin{figure}[t]
			\includegraphics[width=\linewidth]{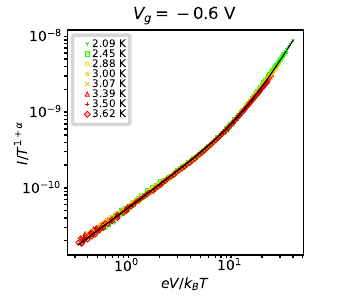}
			\caption{A rescaled current, $\current/T^{1+\alpha}$, as a function of $eV/k_\textrm{B}T$ from the data points in Fig~\ref{Fig:I-Vcurve}.
				The black-solid curve is drawn using Eq.~\eqref{Eq:USC} with the parameters $(\alpha, \gamma, I_0) = (1.3, 0.38, \SI{3.6e-10}{A})$, which were extracted by fitting the data in Fig.~\ref{Fig:I-Vcurve} to Eq.~\eqref{Eq:USC}. 
			} \label{Fig:I-Vscaled}
		\end{figure}

		To this end, we measure $\current$ as a function of the bias voltage $\voltage$ at various temperatures.
		A set of such curves, for top-gate voltage of $V=\SI{-0.6}{V}$, is shown in Fig.~\ref{Fig:I-Vcurve}.
		One can see that the current generally decreases with decreasing temperature $T$, while for a fixed $T$ different slopes for the high-$V$ and low-$V$ regimes can be observed. These features are qualitatively consistent with Eq.~\eqref{Eq:USC}.
		For a fixed top-gate voltage $V_g$, we fit the whole set of $I$-$V$ curves to Eq.~\eqref{Eq:USC} with $I_0$, $\alpha$, and $\gamma$ as the fitting parameters. 
		The rescaled data, together with the fitted curve, are plotted in Fig.~\ref{Fig:I-Vscaled}. 
		We observe that the rescaled data indeed collapse on a single curve, confirming the universal scaling behavior of a TLL.
		
\begin{figure}[t]
			\includegraphics[width=0.95\linewidth]{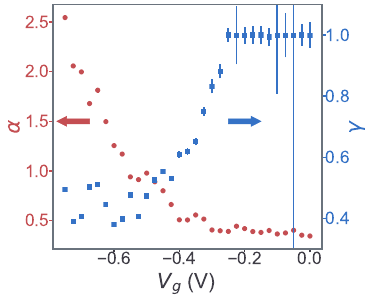}
			\caption{Fitted values of $\alpha$ as a function of the top gate voltage $V_g$ (red circle, left axis).
				The error bars of the fitting are smaller than the markers. 
				Fitted values of $\gamma$ as a function of $V_g$ are also plotted (blue box, right axis).
				For $V_g > \SI{-0.25}{V}$, the fitted values of $\gamma$ are not very reliable, reflected also by larger error bars.
			}\label{Fig:FitParams}
		\end{figure}

		After confirming that the universal scaling holds, and therefore the parameters $\alpha$ and $\gamma$ are reasonably assigned by a fit, we examine their dependence on the carrier density. 
		As the latter is tunable through the top gate voltage, we repeat the above measurements and fittings for various $V_g$ and plot the results in Fig.~\ref{Fig:FitParams}.
		One can see that both parameters change with $V_g$, implying, first of all, that the e-e interaction strength varies with the carrier density.
		To convert the extracted parameters to physical parameters of the system is rather involved and we will do it later.
		Before that, let us comment on the fitted values of $\gamma$ collapsing to 1 for voltages $V_g>\SI{-0.25}{V}$.
		As we already stated, the fitted value of $\gamma$ is determined by the position of the kink in the current-bias curve (for example, in Fig.~\ref{Fig:I-Vscaled} the kink is at $eV/k_\textrm{B}T \approx 10$).
		However, the smaller is $\alpha$ the more straight is the $I$-$V$ curve and the more difficult is to determine the position of the kink from data measured within a fixed voltage range.
		We seem to have less sensitivity to $\gamma$ for $\alpha \lesssim 0.5$, where the fit returns $\gamma=1$. 
		Though error bars become larger here, this value is consistent with the trend observed where $\gamma$ is well detectable.
		Nevertheless, the most interesting part of this plot is on its left end, for large negative values of the top gate $V_g$. 
		Here, $\alpha$ is large, which corresponds to strong e-e interactions, as we will see.
		Also, in the same region, $\gamma$ is around 0.5, corresponding to two tunneling barriers. 
		We are primarily interested in extracting the strength of the e-e interactions in this regime, to see up to where they can be boosted in this platform.
		
\section{Deducing the strength of electron-electron interactions}
		
	\subsection{Description of theoretical methods that we use}	
		We now describe our theoretical analysis, which allows us to extract the strength of the e-e interactions from the observed $\alpha$.
		Motivated by the strong SOIs in InAs, we model each of the wires as a TLL subject to SOIs. 
		To this end, we incorporate the SOI-induced band distortion, which is parametrized by the ratio $\delta v/ v_F$ with $\delta v$ the velocity difference between the two branches of the distorted energy bands~\cite{Moroz1999}; see appendix for details. 
		This band distortion breaks the spin-charge separation of a TLL in Eq.~\eqref{Eq:H_TLL}~\cite{Moroz2000,Moroz2000a} and leads to a coupling between the charge and spin sectors[see Eq.~\eqref{Eq:SOI} in appendix].
		In addition, the SOIs can cause the value of $g_s$ to depart from unity~\cite{Hausler2001,Kainaris2015a}.
		In deriving the current-voltage characteristics, we include both the charge-spin coupling in the Hamiltonian and a general value for the $g_s$ parameter.
		
		The theoretical analysis is complicated not only by the presence of the SOIs, but also by the fact that the conductance depends on the characters and positions of the scatterers (strong or weak, and inside the wire or around its boundary) and also on the value of $\alpha$ itself (larger or smaller than 0.5).
		Including these features is what sets our work apart from preceding studies.
		For the sake of brevity, we delegate the full analysis to Ref.~\cite{Hsu2018b} and state the main results from there in appendix. 
		Here, in the main text, we distill that results further, and only give and comment on the formulas which are used to fit the experimental data.
		
		We start with that, first, we observe $\gamma$ roughly between 1 and 1/2, and, second, that we expect disorder to be generally present in the wires\footnote{As the wires are much longer than the bulk mean free path of \SI{690}{nm}, the disorder (perhaps, in the form of weak potential modulation due to impurities) should play role for the wire resistance.}.
		Justified by these facts, we analyze a system in which {\it two} tunnel barriers and a weak disorder-modulated potential contribute to the wire resistance.
		Precisely, we consider the following scenarios for the two barriers: (A) both of them are in the bulk of a TLL, as illustrated in Fig.~\ref{Fig:g_c}(a), and (B) both of them are around the boundaries of a TLL, between the TLL and a Fermi-liquid lead; see Fig.~\ref{Fig:g_c}(b)
		\footnote{One may consider a third scenario, in which one of barriers is in the bulk and the other is at a boundary. 
		In contrast to our observation, however, it would give different scaling behavior in the high-bias and high-temperature regimes~\cite{Venkataraman2006}. 
		We therefore believe that this scenario is not relevant to our data.}.
		We assume that the contributions from these resistance sources are additive.
		In each scenario, we therefore obtain the total resistance of the wire arising from the tunnel barriers and many weak impurities.			
		Below we first discuss contributions from these sources individually, and then explain how they influence the total wire resistance when they coexist.
		
		In the presence of a single bulk or boundary barrier, we follow Ref.~\cite{Balents1999} to compute the tunnel current through the barrier, which gives the full current-bias curve for various temperatures.
		Then, we generalize our results to the two-barrier case by replacing $V \rightarrow V/2$.
		Further, we use the renormalization-group (RG) method of Refs.~\cite{Kane1992,Giamarchi2003a} to obtain the power-law conductance for a single bulk or boundary barrier (for each barrier type, we again generalize the result for two barriers) in the high-temperature and high-bias limits.
		In these limits, the tunnel conductance derived from the two theoretical approaches 
		(the tunnel-current calculation and the RG method) 
		gives the same exponent.
		In this way, we verified that these approaches give consistent results for the tunnel barriers.
					
		For many weak impurities (that is, disorder potential), on the other hand, the method of Ref.~\cite{Balents1999} is not applicable. 
		We therefore employ the RG method of Refs.~\cite{Kane1992,Giamarchi2003a} to obtain the exponent of the power-law conductance in the high-temperature and high-bias limits. 
		In this case, we take Eq.~\eqref{Eq:USC} as an interpolation formula, with the parameter $\alpha$ replaced by the computed exponent of the power-law conductance and $\gamma = 1$ regardless of the number of impurities.
		
		We now consider the situation with coexisting tunnel barriers and weak impurities, which is relevant to our Scenarios A and B. 
		For generality, we adopt the RG method to determine the power law of the current-voltage dependence for each of the resistance sources.
		Due to the distinct power laws for these sources, we can identify a single term which dominates the resistance (in the RG sense) for a given strength of e-e interactions.
		In each scenario, neglecting the subdominant term, we obtain the expression for $\alpha$ in terms of the intrinsic interaction parameters $g_{\rm c}$ and $g_{\rm s}$, as well as the ratio $\delta v/ v_F$.
		
		Finally, in order to convert the observed power $\alpha$ to the value of $g_{\rm c}$, we need the values of the SOI-induced parameters, $\delta v/ v_F$, and the departure of $g_s$ from 1.
		For parameters relevant to our experiment, we estimate $\delta v/ v_F \lesssim 0.1$.
		As $1-g_s$ scales with the same quantity, $\delta v/ v_F$~\cite{Kainaris2015a}, we find that the modification in $g_s$ is also negligible~\cite{Hausler2001}. 
		These findings mean that we can use the zero-SOI expressions for $\alpha$.
		In addition, when the wires are close to being depleted, which is the strong-interaction regime of our primary interest, $\delta v/ v_F$ becomes vanishingly small, making our approximations even more accurate.
		We note that, even with these approximations, the conversion from the observed $\alpha$ to $g_c$ is still complicated due to various types of the resistance sources.
		In the following, we present the derived expression of $\alpha$ and its approximated form for Scenarios A and B.
		Utilizing the latter form, we extract the $g_{\rm c}$ value from the observed $\alpha$.
				
	\subsection{Conversion of $\alpha$ to $g_c$ in Scenario A (two bulk tunnel barriers and weak disorder)}
\begin{figure}[t]
			\includegraphics[width=0.95\linewidth]{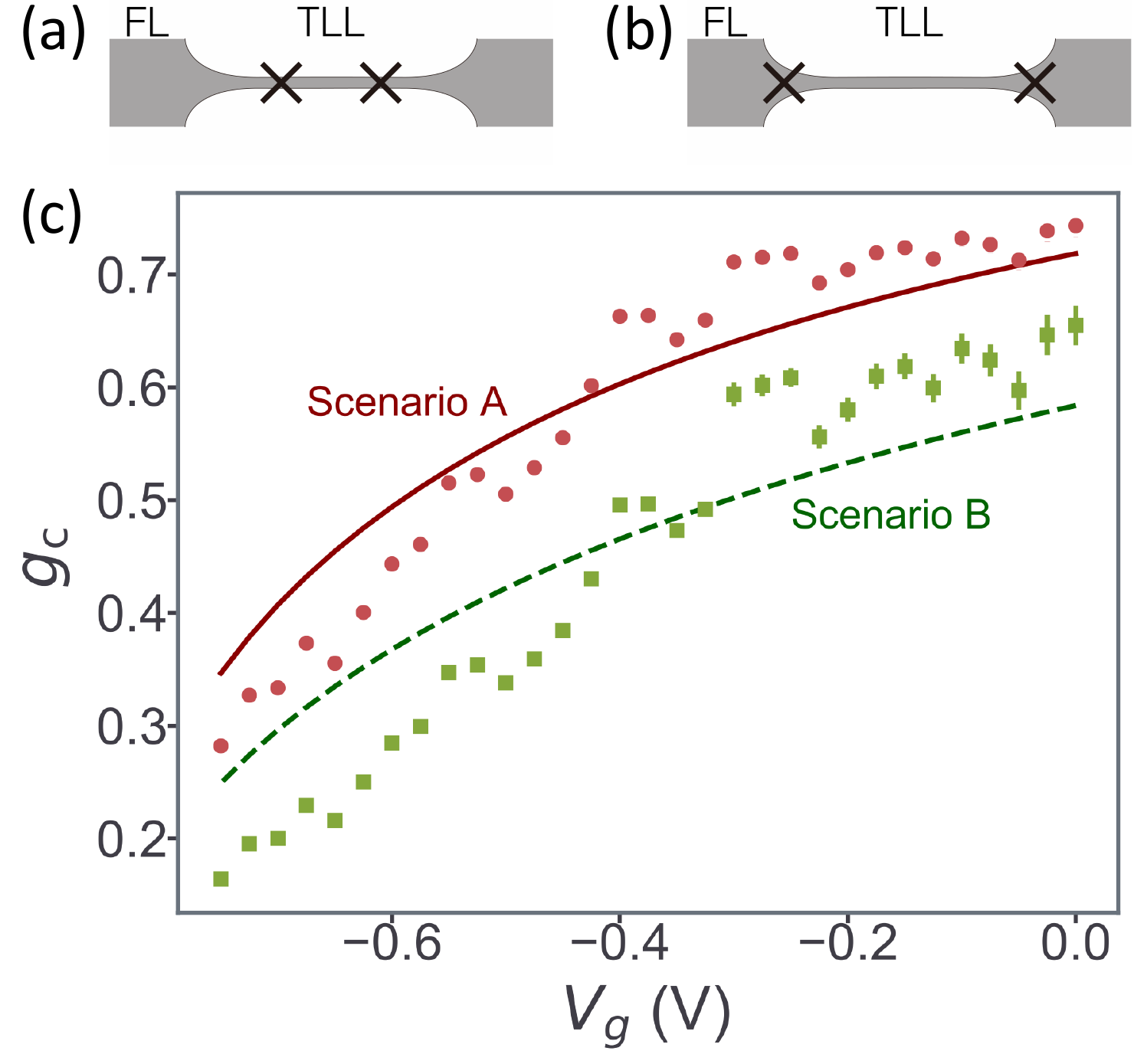}
			\caption{(a) A schematic illustration of Scenario A: there are two bulk barriers, each acting as a TLL-TLL junction, and many weak impurities (not shown). 
				(b) A schematic illustration of Scenario B:  there are two boundary barriers, each acting as a TLL-Fermi liquid (FL) junction, and many weak impurities. 
				(c) Extracted values of the interaction parameter $g_\textrm{c}$ as a function of the top gate voltage $V_g$ for the two scenarios. 
				The red-solid and green-dashed curves are the fits to Eq.~\eqref{Eq:gcth}, with the fitting parameter $w$ (the wire transverse size) \SI{87}{nm} and \SI{47}{nm}, respectively.
			}\label{Fig:g_c}
		\end{figure}

		In Scenario A, the tunnel current is given by Eq.~\eqref{Eq:USC} with $\gamma=1/2$ and $\alpha$ replaced by[appendix]
		\begin{subequations}\label{Eq:gcTTwSOI}
			\begin{align}
				\ab &= \left(
					\frac{1}{g_\textrm{c}^\prime} +\frac{1}{g_\textrm{s}^\prime}
				\right)(\cos^2 \theta+g_0^2 \sin^2 \theta)-2\\
				&\approx \frac{1}{g_c}-1,
			\end{align}
		\end{subequations} 
		where the approximation is valid for parameters relevant here. 
		In the above, $\theta$ is a small parameter characterizing the strength of the SOIs, and the explicit forms of $g^\prime_\nu$, $g_0$, and $\theta$ are given in appendix.
		On the other hand, the interpolation formula for weak impurities is given by Eq.~\eqref{Eq:USC}, with $\gamma=1$ and $\alpha$ replaced by
		\begin{subequations}\label{Eq:weakimp}		
			\begin{align}
				\ai &= 2 - \cos^2 \theta (g_\textrm{c}^{\prime} + g_\textrm{s}^{\prime} ) - g_0^2 \sin^2 \theta \left( \frac{1}{g_\textrm{c}^{\prime}} + \frac{1}{g_\textrm{s}^{\prime}} \right)\\
				&\approx 1-g_\textrm{c},
			\end{align}
		\end{subequations}
		where, again, the second line stems from the approximation valid for relevant parameters.
		Importantly, for any repulsive interaction $g_{\rm c} \le 1$, the approximated value is bounded $\ai \le 1$, allowing
		us to rule out the weak impurities as the source of the observed value $\alpha >1$ in the low-$V_{g}$ regime.
		Further, for any $g_\textrm{c}<1$, 
		one has $1/g_c-1 > 1-g_c$, so that the resistance from the bulk tunnel barriers dominates the one from weak impurities. 
		We therefore assign the observed power law to bulk barriers and use 
		\begin{align}
			\aA  &=\frac{1}{g_\textrm{c}}-1,
				\label{Eq:gcTT}
		\end{align}
		to extract the $g_\textrm{c}$ values from the data in Fig.~\ref{Fig:FitParams}.
		In Fig.~\ref{Fig:g_c}(c), we plot the extracted $g_\textrm{c}$ value as a function of $V_g$.
		The lowest value $g_\textrm{c}=0.28$ corresponds to very strong e-e interactions in a wire with low electron density.
		
		To further check the consistency of our procedure, we fit the extracted values for $g_\textrm{c}$ to the formula\cite{Maciejko2009,Kane1992}
		\begin{align}
			g_\textrm{c} &= 
			\left[
				1+\frac{e^2}{\pi ^2 \varepsilon \hbar v_\textrm{F}} \ln \left( \frac{D^2}{d w}\right)
			\right]^{-1/2}.
			\label{Eq:gcth}
		\end{align}
		In this equation, derived by estimating the compressibility of the electron gas with Coulomb interactions screened by a conducting plane (the top gate), $D$ is the distance between the wire and the top gate, $d$ is the quantum well thickness, $w$ is the wire width, and $\varepsilon$ is the dielectric constant.
		For our device, we have $D=\SI{300}{nm}$, $d=7$ nm, and $\varepsilon =15.15\,\varepsilon_0$\cite{Levinshtein1996}.
		Using $w$ as a fitting parameter, we get the curve shown in Fig.~\ref{Fig:g_c}(c), showing a good correspondence with $g_c$ fitted from the data.
		Further, the fitted value $w=\SI{87}{nm}$ is consistent with the nominal width of $\SI{100}{nm}$.
		Given $w$, we estimate the wire subbands level spacing $\frac{\hbar^2}{2m^*}\left(\frac{2\pi}{w}\right)^2 \approx \SI{8.64}{m\electronvolt}$%
			\footnote{From the stacking structure of the wafer, we estimate the Fermi energy $E_\textrm{F}=1.13\times 10^2\times [V_g-(-0.86)]^2$ [\si{m\electronvolt}] and the Fermi velocity $v_\textrm{F}=1.31\times 10^{13}\times [V_g-(-0.86)]w$ [\si{m/s}][See appendix]. Here, $V_g$ is in units of \si{V}, $w$ is in units of \si{m}.}, %
		corresponding to $E_\textrm{F}$ at $V_g = \SI{-0.54}{V}$.
		The estimate therefore suggests that our device is in the single-channel regime for $V_g<\SI{-0.54}{V}$, where an approximately constant value of $\gamma=1/2$ is seen on Fig.~\ref{Fig:FitParams}.%
			\footnote{In addition, the level spacing is large enough so we can ignore higher subbands at temperatures of our measurements. In Ref.~\cite{Heedt2016}, the subband level spacing in a \SI{100}{nm}-diameter InAs nanowire with isotropic cross-section of about \SI{8}{m\electronvolt} was found, similar to our estimate here.}% 
		Given all these cross-checks, we conclude that Scenario A provides a consistent interpretation of the measured data.
		
	\subsection{Conversion of $\alpha$ to $g_c$ in Scenario B (two boundary tunnel barriers and weak disorder)}
		
		We now consider Scenario B, in which the tunnel current through the boundary barriers is given by Eq.\eqref{Eq:USC} with $\gamma=1/2$ and[appendix]
		\begin{subequations}
			\begin{align}
				\aed &=\frac{1}{2}
				\left(
					\frac{1}{g_\textrm{c}^\prime}+\frac{1}{g_\textrm{s}^\prime}
				\right)(\cos^2 \theta+g_0^2 \sin^2 \theta)-1\\
				&\approx\frac{1}{2g_\textrm{c}}-\frac{1}{2}.
				\label{Eq:aed_noSOI}
			\end{align}
		\end{subequations}
		The contribution of weak impurities is the same as in Scenario A, characterized by $\gamma=1$ and $\ai \approx 1-g_\textrm{c}$.
		In contrast to Scenario A, now weak impurities become dominant over boundary tunnel barriers for $\alpha<0.5$.
		The observed parameter $\alpha$ is therefore related to the interaction parameter $g_\textrm{c}$ through
		\begin{align}
			\aB &\approx 
			\left\{
				\begin{array}{ll}
					\frac{1}{2g_\textrm{c}}- \frac{1}{2}, &	\mbox{for $\alpha \ge 0.5$ (barriers);} \\
					1-g_\textrm{c},	&	\mbox{for $\alpha \le 0.5$ (impurities).}
				\end{array}
			\right.
			\label{Eq:gcTF}
		\end{align}       		
		The extracted $g_\textrm{c}$ is shown in Fig.~\ref{Fig:g_c}(c) along with the fit to Eq.~\eqref{Eq:gcth}.
		Here, the $g_\textrm{c}$ values are even smaller and reach as low as 0.16.
		
		Next, we discuss how additional features of the extracted values for $\alpha$ and $\gamma$ fit with the assumptions of Scenario B.
		Namely, as $V_g$ decreases, the extracted $\gamma$ value decreases from unity and approaches 0.5 around $V_g = \SI{-0.4}{V}$.
		At the same voltage, $\alpha$ steps across 0.5, which is the transition point between the two expressions in Eq.~\eqref{Eq:gcTF}.
		Such a feature can be well captured by Scenario B, where $\gamma$ should be unity when weak impurities dominate and $1/2$ when the boundary tunnel barriers dominate.
		The fit to Eq.~\eqref{Eq:gcth} gives a value $w=\SI{47}{nm}$ and the associated subband level spacing of \SI{29.6}{m\electronvolt}, indicating that the wire is in the single-channel regime for $V_g < \SI{-0.23}{V}$, where the extracted $\gamma$ drops below 1.
		Thus, we conclude that Scenario B is also in agreement with several aspects of the data.

	\subsection{Conclusion on the considered scenarios}		

		Both Scenario A and B are reasonable and it is difficult to decide for one. 
		Scenario A gives somewhat better agreement with Eq.~\eqref{Eq:gcth}; however, we do not deem a quantitative discrepancy to such a simple theory as very informative. Arguably, the weakest point of Scenario A is the assumption that each wire contains exactly two tunnel barriers in its interior%
			\footnote{If they originate in random disorder, there is no reason for such uniformity. 
				On the other hand, one could argue that disorder average over many parallel wires might result in a scaling curve with some effective number of tunnel barriers, being here close to 2. 
				However, performing such a calculation would require to adopt some ad-hoc assumptions about the statistical distributions of the strength and position of the tunnel barriers. We therefore do not follow this idea.
			}.
		On the contrary, in Scenario B the tunnel barriers are formed at the wire ends and having two per wire is natural. 
		Nevertheless, we emphasize that regardless of which scenario is realized, both support our main conclusion that strong and gate-tunable e-e interactions are present in the wires.
		
%%%%%%%%%%%%%%%%%%%%%%%%%%%%%%%%%%%%%%%%%%%%%%%%%%%%%

	\begin{table*}[t]
		\caption{
			Deduced interaction parameters ($g_{\rm c}$) of one-dimensional systems as reported in experiments, including the present work (shaded row).
			The description of the entries is as follows.
			The first column gives the material(s) used in the listed references. 
			The second column lists the extracted $\alpha$ parameter (if available) from the observed quantity given in the sixth column.
			Based on the resistance sources attributed in the references, the corresponding parameters $\ab$, $\aed$ and $\ai$ are given (we label those with unspecified sources with an unsubscripted $\alpha$).
			The third and fourth columns list the interaction parameter $g_\textrm{c}$ either quoted from the references (in black) or deduced from the $\alpha$ value (in red) using Table~\ref{Tab:tunnel}.
			The third column includes the $g_{\rm c}$ value deduced from $\ab$ or those with unknown sources. 
			The fourth column includes those from either $\aed$ or $\ai$.
			For $\alpha$ value with unknown resistance sources, we deduce $g_{\rm c}$ values for all impurity types considered here.
			The extracted $\gamma$ value (if available) is given in the fifth column.
			The notations $G$, $T$, and $R$ denote the conductance, temperature, and resistance, respectively. 
			The abbreviations NW, CNT, VG, and CE stand for nanowire, carbon nanotube, V-groove, and cleaved edge, respectively.}
		\begin{tabular}{c|c|c|c|c|c}
			material [Ref] & \multicolumn{1}{c|}{extracted $\alpha$}  & \multicolumn{1}{c|}{$g_\textrm{c}$ deduced\footnote{Here we use $\ab = 1/g_{\rm c} - 1$ for NWs and $\ab = (1/g_{\rm c} - 1)/2$ for CNTs. Note that here we intentionally use the same notation $g_{\rm c}$ for both NWs and CNTs; see Table~\ref{Tab:tunnel} for general expressions.}} & \multicolumn{1}{c|}{$g_\textrm{c}$ deduced\footnote{Here we use $\aed = (1/g_{\rm c} - 1)/2$ and $\ai= 1- g_{\rm c}$ for NWs, and $\aed = (1/g_{\rm c} - 1)/4$ and $\ai = (1 - g _{\rm c})/2$ for CNTs; see Table~\ref{Tab:tunnel} for details.}}  & \multicolumn{1}{c|}{$\gamma$} & \multicolumn{1}{c}{observed}  \\ % multicolumn: for centering 
			& \multicolumn{1}{c|}{from experiment} & \multicolumn{1}{c|}{from $\ab$} & \multicolumn{1}{c|}{from $\aed$~or~$\ai$} & & \multicolumn{1}{c}{quantity}   \\ % multicolumn: for centering
			\hhline{=|=|=|=|=|=}
			MoSe NW~\cite{Venkataraman2006} & $\ab=$0.61--6.6; $\aed=$0.94--5.2  & \red{0.13--0.62} & \red{0.09--0.35}~($\aed$) & 0.25%
				~\footnote{In this reference, while the universal scaling behavior was observed and thus the $\gamma$ value was obtained, the value for $\alpha$ was extracted from the power-law conductance rather than the from full current-voltage curve.
				} %
			& $G \propto T^\alpha$  \\ %
			\cellcolor[gray]{0.8} InAs NW%
			& \cellcolor[gray]{0.8} $\alpha=$ 0.35--2.5 & \cellcolor[gray]{0.8} \red{0.28--0.74} & \cellcolor[gray]{0.8} \red{0.16--0.65} (both) & \cellcolor[gray]{0.8} 0.5--1.0 & \cellcolor[gray]{0.8} Eq.~\eqref{Eq:USC} \\ %
			Multi-wall CNT~\cite{Graugnard2001} &  $\aed =$ 0.36--0.95 & -- & 0.21--0.41~($\aed$)  & 0.05--0.24 & $G$~\footnote{On top of the universal scaling conductance, additional phenomenological parameters are required for their fitting.} %
			 \\ %
			InAs NW\footnote{In this reference, the device forms a quantum dot.}~\cite{Hevroni2016} & -- & \red{0.23}%
				~\footnote{This reference reported a small value 0.38 for the effective interaction parameter $g=(1/2g_\textrm{c}+1/2g_\textrm{s})^{-1}$, which was attributed to $g_\textrm{s}<1$ due to the SOIs. 
				In contrast, our work indicates that the effects of the SOIs on the interaction parameter are negligible for relevant strength of the SOIs.
				With the assumption $g_\textrm{s}=1$, the value of $g_\textrm{c}$ in this reference becomes 0.23. We use the latter value for the table entry here.}% 
			& -- & -- & $G_\textrm{max} \propto T^{\frac{1}{g}-2}$~\footnote{The notation $G_\textrm{max}$ denotes the conductance value of the Coulomb peak.}  \\ %
			Single-wall CNT~\cite{Postma2000} &  $\ab=$ 1.4%
		        & 0.26 & -- & 0.6~\footnotemark[3] & $G \propto T^\alpha$  \\ %
			Multi-wall CNT~\cite{Bachtold2001} & $\ab=$ 1.24; $\aed =$ 0.6~\footnote{In this reference, the tunnel conductance from a FL lead into {\it the bulk of a TLL} is also measured. It leads to a different power law, whose exponent is, however, not included here.} & \red{0.29} & \red{0.29}~($\aed$)  & -- & $G \propto T^\alpha$  \\ %
			Single-wall CNT~\cite{Bockrath1999} & $\aed=$ 0.6~\footnotemark[8] & -- & \red{0.29}~($\aed$) & 0.46--0.63~\footnotemark[3] & $G \propto T^\alpha$  \\%
			NbSe$_3$ NW~\cite{Slot2004}   & $\ab=$ 2.15--2.2 & \red{0.31--0.32} & -- & $\frac{1}{100}$--$\frac{1}{77}$~\footnotemark[3] 
			& $R \propto T^{-\alpha}$  \\ %
			GaAs VG~\cite{Levy2006} & -- &  --  & 0.45--0.66~($\ai$) & -- & $\delta G_1$~\footnote{The notation $\delta G_1$ denotes the conductance correction of the first conductance plateau.} \\ %
			GaAs/AlGaAs CE~\cite{Rother2000} & $\ai=$ 0.5 & -- & \red{0.50}~($\ai$) & -- & $\delta G_1$ \\ %
			GaAs VG~\cite{Levy2012} &  -- & 0.54--0.66 & -- & -- & $G \propto T^{\frac{1}{g_\textrm{c}}-1}$  \\ %
			Single-wall CNT~\cite{Lee2004} & -- & 0.55 & -- & -- & STM imaging  \\ %
			GaAs/AlGaAs~\cite{Asayama2002} &-- & 0.6 & -- & -- & $\Delta R_\textrm{bs}$~\footnote{The notation $\Delta R_\textrm{bs}$ denotes the backscattering resistance due to Bragg reflection.}  \\ %
			GaAs/AlGaAs~\cite{Tarucha1995}  & -- &  -- & 0.65--0.7~($\ai$) & -- & $\delta G_1$ \\ %
			GaAs/AlGaAs CE~\cite{Auslaender2000} & -- & 0.66--0.82 & -- & -- & $\Gamma_i \propto T^{\frac{1}{g_\textrm{c}}-1}$~\footnote{The notation $\Gamma_i$ denotes the full width at half maximum of a Coulomb peak.}\\ %
			GaAs NW\footnote{In this reference, a core-shell nanowire was used.}~\cite{Liu2001} & $\alpha=$ 0.02--0.23 & \red{0.81--0.98} & \red{0.77--0.98}~($\ai$)\footnote{Alternatively, assuming that disorder is absent within the wire, the $g_{\rm c}$ value deduced from $\aed$ follows as 0.68--0.96.} & -- & $G \propto T^\alpha$ \\ %
			Multi-wall CNT~\cite{Lucot2011} & $\alpha=$ 0.02--0.05 & \red{0.91--0.96} & \red{0.90--0.96}~($\ai$)\footnote{Alternatively, assuming that disorder is absent within the nanotube, the $g_{\rm c}$ value deduced from $\aed$ follows as 0.83--0.93.} & -- & $G \propto T^\alpha$  \\ %
		\end{tabular}	
		\label{Tab:gc}
	\end{table*}

%%%%%%%%%%%%%%%%%%%%%%%%%%%%%%%%%%%%%%%%%%%%%%%%%%%%%

\section{Comparison to e-e interaction strengths reported in literature}
		Before concluding, we compare the e-e interaction strength found here with previous experiments. To make sensible comparison of numerous references, we convert---whenever possible---to unified parameters, being $g_\textrm{c}$ and $\alpha$ in the notation of this article. 
		We include one-dimensional systems regardless of materials or measurement types and arrive at Table~\ref{Tab:gc}, with entries ordered by the lowest value of $g_c$ achieved in a given reference.
		In general, systems with well-defined single channels (e.g. single-wall carbon nanotubes) tend to have smaller values of $g_\textrm{c}$ (stronger interactions\footnote{In the TLL model that we work with here, the constants $g$ are the only parameters defining the strength of the electron-electron interactions. The value of the Fermi velocity, or the relation of the kinetic to interaction energies, would also need to be considered to judge the ``strength'' of the interactions in a broader context. Here we do not consider such implications and mean the statements on the e-e interaction strength as solely the statements on the value of constants $g$.}) due to suppression of scattering and stronger spatial confinement.
		A smaller mass of InAs compared to GaAs is also beneficial for a smaller $g_\textrm{c}$, giving a larger level spacing and a well-defined single channel.

\section{Conclusions}
		To conclude, we investigate quantum wires etched from an InAs quantum well and find that they possess strong e-e interactions.
		This finding is based on observation of universal scaling of the current as a function of the bias voltage and temperature, from which the Luttinger liquid interaction parameter can be fitted. 
		The fitting requires a theory for the conversion of the observed exponent $\alpha$ of the power law dependence of the conductance to the e-e interaction strength parameter $g_{\rm c}$ in the TLL Hamiltonian. 
		The relation between $\alpha$ and $g_{\rm c}$ depends on the character and positions of the scatterers as sources of resistance. 
		For the case of finite SOIs, we provide the main results of such conductance theory here. Its most important conclusion is that for strong e-e interactions, the effects of the SOIs on the relation between $\alpha$ and $g_{\rm c}$ are negligible. 
		It reassures us that the large values of $\alpha$ that we observe are due to genuinely strong e-e interactions, and not, for example, an artifact of strong SOIs. 
		All together, our work demonstrates that an etched InAs quantum wire is a promising platform offering quasi one-dimensional channel with strong SOIs, and strong and gate-tunable e-e interactions.
		
\section*{Acknowledgments}
		This work was partially supported by a Grant-in-Aid for Young Scientific Research (A) (Grant No.JP15H05407), Grant-in-Aid for Scientific Research (B)  (No.JP18H01813), Grant-in-Aid for Scientific Research (A) (Grant No.JP16H02204), Grant-in-Aid for Scientific Research (S) (Grant No.JP26220710), JSPS Research Fellowship for Young Scientists (Grant No.JP18J14172), Grants-in-Aid for Scientific Research on Innovative Area `Nano Spin Conversion Science' (Grants No.JP17H05177) and a Grant-in-Aid for Scientific Research on Innovative Area `Topological Materials Science' (Grant No.JP16H00984) from MEXT, JST CREST(Grant No.JPMJCR15N2), the ImPACT Program of Council for Science,Technology and Innovation (Cabinet Office,Government of Japan).
		HK acknowledges support from RIKEN Incentive Research Projects, and JSPS Early-Career Scientists (Grant No.JP18K13486).
		The authors would like to thank Michihisa Yamamoto and Ivan V. Borzenets for support of experiment and measurement equipment.

%\clearpage
\appendix

\section{Experimental details}
	\subsection{Chemical etching and crystal axis}
\begin{figure}[h]
		\includegraphics[width=0.95\linewidth]{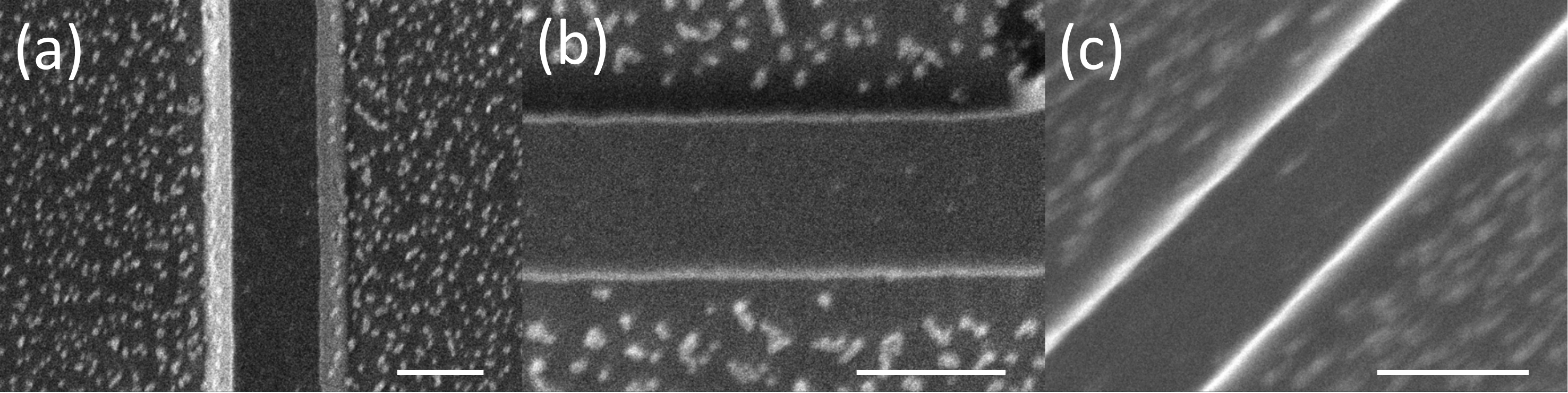}
		\caption{SEM images of quantum wires chemically etched out from the wafer grown along [001] direction.
		The wires are along (a)[1$\bar{1}$0], (b)[110], (c)[010] direction, respectively.
		Scale bars in each figure indicates length of \SI{500}{nm}.
		From the brightness of the edges compared to dots surrounding the wires, each wire's cross-section is inferred as a trapezium, a reverse trapezium and a rectangle, respectively.}\label{Fig:SEM}
	\end{figure}

		To form our Hall-bar and quantum-wire devices, we use chemical etching by diluted H$_2$O$_2$ and H$_2$SO$_4$.
		It is well known that the rates of etching speed depend on the crystal axis of the samples, consequently so do the edge shapes of the devices.
		We tested the etching process on a trial wafer and confirmed such dependency by SEM.
		Figure~\ref{Fig:SEM} shows SEM images of wires formed in (a)[1$\bar{1}$0], (b)[110], (c)[010] direction.
		The SEM image reflects the slope of edges, and therefore it enables us to identify cross-sections of these wires as a trapezium, a reverse trapezium and a rectangle, respectively.
		Based on these findings, we choose to measure on wires formed along [010] direction so that we can determine the width of the quantum wires more precisely, being the same as the width of their top-surface.

	\subsection{Estimation of the gate dependence of electronic density in wires}
		From the stacking structure of the quantum well and \SI{260}{nm}-thick cross-linked polymethyl methacrylate (PMMA), we estimate the top gate capacitance of \SI{2.71e-16}{F}.
		We take the dielectric constants of PMMA and InAlAs to be 4 and 13.59, respectively\cite{Li2015a,Littlejohn1993}.
		With this, we estimate the Fermi energy $E_\textrm{F}=1.13\times 10^2\times [V_g-(-0.86)]^2$ [meV] and the carrier density $n=8.31\times 10^5\times [V_g-(-0.86)]$ [cm$^{-1}$], respectively.
		Owing to the high mobility of the quantum well, longer uniform quantum wires can be realized compared to self-assembled nanowires\cite{Hevroni2016}.

\section{Theoretical analysis}
		In this appendix we present the main results of our theoretical analysis. 
		We first discuss the effects of the spin-orbit interactions (SOIs).
		We will then discuss the formulas which we use to extract the interaction parameters in various scenarios.
		In addition, we summarize the expressions of the exponent $\alpha$ for various Tomonaga--Luttinger liquid (TLL) systems in existing literature.
		
	\subsection{Effects of SOIs}
		In this section we discuss the effects of SOIs on the power-law conductance and current-voltage curve of a TLL.
		The motivation for this calculation is to examine how the SOIs affect the observed parameter $\alpha$.
		Namely, whereas the observed universal scaling behavior in the current-voltage characteristics unambiguously establishes the TLL behavior of our quantum wires, it remains to be clarified whether the rather large $\alpha$ value (implying small extracted $g_\textrm{c}$) in the low-density regime is not an illusion owing to the strong SOIs in InAs.
		
		First of all, we remark that it is known that in the absence of a magnetic field, the SOIs can be gauged away in strictly one dimensional system, thereby having no influence on observable quantities~\cite{Braunecker2010,Meng2014,Kainaris2015a}.
		In a quasi-one-dimensional geometry such as the etched quantum wires in our experiments, however, the interplay between the SOIs and the finite transverse confinement potential that defines the width of a quantum wire can modify the band structure, leading to different velocities for different branches in the spectrum~\cite{Moroz1999,Meng2014}.
		It was shown that such an effect destroys the spin-charge separation~\cite{Moroz2000,Moroz2000a}, leading to a coupling between the spin and charge sectors in Eq.~\eqref{Eq:H_TLL} in the main text.
		
		To investigate whether such a coupling alters the observed $\alpha$ value, we theoretically analyze its effects on the current-voltage characteristics.
		In the following we first outline our calculation based on the TLL formalism, and then give our results on various types of the impurities.
		To be specific, we consider the impurities which are either strong or weak (acting as tunnel barriers or potential disorder), and for the former type we further consider whether they locate in the bulk or at the boundaries (ends) of the wires.
		
		Before continuing, let us comment on possible origins of the tunnel barriers at the boundaries of the wire.
		We first clarify that these ``boundary barriers'' may be located close to, but not exactly at the physical boundary between the wire and a lead.
		In fact, as discussed in Ref.~\cite{Balents1999}, a barrier can be considered as a boundary barrier if the distance from the wire boundary is shorter than the scales $\hbar v_F/(k_B T)$ and $\hbar v_F/(eV)$.
		Since for our experiments these length scales are typically of order $O$(100~nm)--$O$(1~$\mu$m), comparable to the bulk mean free path $\sim$ 690~nm (regarded as the average impurity separation), the tunnel barriers may arise from randomly distributed impurities near the boundaries.
		In addition, a poor contact between the wire and a lead may also be regarded as a boundary barrier. 
		Therefore, the ``boundary barrier'' discussed here may come from either impurities or contacts with poor transmission in the vicinity of (but not necessarily at) the boundaries (ends) of the wire.
		
		\begin{widetext}
		To proceed, we follow Refs.~\cite{Moroz2000,Moroz2000a} and add the following term to Eq.~\eqref{Eq:H_TLL} in the main text, 
		\begin{align}
			H_{\textrm{so}} &= \delta v \int \frac{\hbar dx}{2\pi} \left\{ \left[ \partial_x \phi_\textrm{c}(x)\right] \left[\partial_x \theta_\textrm{s}(x) \right] + \left[ \partial_x \phi_\textrm{s}(x) \right] \left[\partial_x \theta_\textrm{c}(x) \right] \right\}.
			\label{Eq:SOI}
		\end{align}
		It reflects the presence of SOIs as a velocity difference $\delta v$ between the two branches of the energy spectrum.
		Since the full Hamiltonian $H_{\textrm{TLL}}+H_{\textrm{so}}$ is still quadratic in the bosonic fields, we can diagonalize it to get
		\begin{subequations} 
			\begin{align}
				H_{\textrm{TLL}}^{\prime} &\equiv H_{\textrm{TLL}} + H_{\textrm{so}} 
				= \sum_{\nu = c, s} \int \frac{\hbar dx}{2\pi} \left\{
				u_{\nu}^{\prime} g_{\nu}^{\prime} \left[ \partial_r \theta_{\nu}^{\prime}(x) \right]^2 + \frac{u_{\nu}^{\prime}}{g_{\nu}^{\prime}} \left[ \partial_x \phi_{\nu}^{\prime}(x) \right]^2 \right\},
				\label{Eq:TLL}
			\end{align}
			where the modified TLL parameters and velocities are given by
			\begin{align}
				g_\textrm{c}^{\prime} = & \frac{g_\textrm{c} g_0}{g_\textrm{s}} \left[ \frac{(g_0^2 +g_\textrm{s}^2) +(g_\textrm{s}^2 -g_0^2) \cos (2\theta) + g_0 g_\textrm{s}^2 \frac{\delta v}{v_F} \sin (2\theta) } {(g_0^2 +g_\textrm{c}^2) +(g_0^2 -g_\textrm{c}^2) \cos (2\theta) + g_0 g_\textrm{c}^2 \frac{\delta v}{v_F} \sin (2\theta)}\right]^{1/2}, \\
				g_\textrm{s}^{\prime} = & \frac{g_\textrm{s} g_0}{g_\textrm{c}} \left[ \frac{(g_0^2 +g_\textrm{c}^2) +(g_\textrm{c}^2 -g_0^2) \cos (2\theta) - g_0 g_\textrm{c}^2 \frac{\delta v}{v_F} \sin (2\theta) } {(g_0^2 +g_\textrm{s}^2) +(g_0^2 -g_\textrm{s}^2) \cos (2\theta) - g_0 g_\textrm{s}^2 \frac{\delta v}{v_F} \sin (2\theta)}\right]^{1/2}, \\
				u_\textrm{c}^{\prime} = & \frac{v_F}{2g_0 g_\textrm{c} g_\textrm{s}} \left[(g_0^2 +g_\textrm{c}^2) +(g_0^2 -g_\textrm{c}^2) \cos (2\theta) + g_0 g_\textrm{c}^2 \frac{\delta v}{v_F} \sin (2\theta) \right]^{1/2} \nonumber \\
				& \times \left[ (g_0^2 +g_\textrm{s}^2) +(g_\textrm{s}^2 -g_0^2) \cos (2\theta) + g_0 g_\textrm{s}^2 \frac{\delta v}{v_F} \sin (2\theta) \right]^{1/2}, \\
				u_\textrm{s}^{\prime} = & \frac{v_F}{2g_0 g_\textrm{c} g_\textrm{s}} \left[(g_0^2 +g_\textrm{s}^2) +(g_0^2 -g_\textrm{s}^2) \cos (2\theta) - g_0 g_\textrm{s}^2 \frac{\delta v}{v_F} \sin (2\theta) \right]^{1/2} \nonumber \\
				& \times \left[ (g_0^2 +g_\textrm{c}^2) +(g_\textrm{c}^2 -g_0^2) \cos (2\theta) - g_0 g_\textrm{c}^2 \frac{\delta v}{v_F} \sin (2\theta) \right]^{1/2},
			\end{align}
			with the parameters
			\begin{align}
				g_0 = & \frac{\sqrt{2} g_\textrm{c} g_\textrm{s} }{ \sqrt{g_\textrm{c}^2 + g_\textrm{s} ^2}}, \\
				\theta = & \frac{1}{2} \arctan \left( \frac{\delta v }{v_F}  \frac{ \sqrt{2} g_\textrm{c} g_\textrm{s} \sqrt{g_\textrm{c}^2 + g_\textrm{s}^2} }{ g_\textrm{s}^2 - g_\textrm{c}^2} \right).
			\end{align}
		\end{subequations}
		In the absence of the SOIs, we have $\delta v, \theta \rightarrow 0$, and therefore $(g_\textrm{c}^{\prime}, g_\textrm{s}^{\prime}, u_\textrm{c}^{\prime}, u_\textrm{s}^{\prime}) \rightarrow (g_\textrm{c}, g_\textrm{s}, u_\textrm{c}, u_\textrm{s})$.
		Using the model in Ref.~\cite{Moroz1999}, we have estimated that for the parameters relevant to our experiments, the value of $\delta v /v_F$ is at most around $0.1$ and becomes vanishingly small when the system is close to being depleted. 
		We remark that Ref.~\cite{Moroz2000} obtains a similar estimated value $\delta v/v_F \approx 0.1$--0.2.
		\end{widetext}
		
		With the diagonalized Hamiltonian Eq.~\eqref{Eq:TLL}, we are able to compute the tunnel current and the conductance of the quantum wires.
		Leaving the details for a separate publication~\cite{Hsu2018b}, here we state our results and discuss their relevance to our experiment.
		
		As mentioned in the main text, we consider several scenarios in which different types of the impurities are present.
		We first consider a single strong impurity in the bulk.
		By modeling the impurity as a tunnel barrier, which corresponds to a TLL-TLL junction, we compute the tunnel current through the barrier. 
		For relevant strength of SOIs, we obtain Eq.~\eqref{Eq:USC} in the main text, with the parameters $\gamma=1$ and $\alpha$ replaced by
		\begin{align}
			\ab (g_\textrm{c}^{\prime},g_\textrm{s}^{\prime},\theta) &= \left( \frac{1}{g_\textrm{c}^{\prime}} + \frac{1}{g_\textrm{s}^{\prime}} \right) \left( \cos^2 \theta + g_0^2 \sin^2 \theta \right) -2,
			\label{Eq:alpha_bulk}
		\end{align}
		where the arguments $(g_\textrm{c}^{\prime},g_\textrm{s}^{\prime},\theta)$ are themselves functions of $(g_\textrm{c},g_\textrm{s},\delta v)$.
		The exponent Eq.~\eqref{Eq:alpha_bulk} is given in Eq.~\eqref{Eq:gcTTwSOI} in the main text.
		In the presence of several bulk barriers,  we assume that they induce comparable resistance. The tunnel current through the wire is then given by Eq.~\eqref{Eq:USC} with the same $\ab$ as Eq.~\eqref{Eq:alpha_bulk} and with $\gamma $ equal to the inverse of the barrier number.
		
		An alternative approach based on the renormalization-group tools~\cite{Kane1992,Giamarchi2003a} can be employed to compute the power-law conductance in the high-temperature ($k_B T \gg eV$) and high-bias ($eV \gg k_B T$) limits.
		In the presence of a single bulk barrier, the power-law conductance can be summarized as
			\begin{align}
				G_{\rm bulk} (T,V) \propto& \,{\rm Max}(k_B T, eV)^{\ab}, 
			\end{align}
		which is characterized by the same parameter $\ab$.
		Similar to the tunnel current, the above formula can be generalized for several bulk barriers upon replacing $V \rightarrow \gamma V$ with $1/\gamma$ being the barrier number.
		It can be shown that $G_{\rm bulk} (T,V) $ is consistent with the current-voltage characteristics [Eq.~\eqref{Eq:USC} in the main text] in the high-temperature and high-bias limits. 
		It demonstrates the compatibility of the two approaches.
		
		We now analyze how the SOIs influence the current-voltage characteristics through the parameter $\ab$.
		It is useful to define an effective interaction parameter $g_{\rm c,eff}$, such that all the effects of $\delta v/v_F$ are incorporated into a single parameter. 
		To be specific, we define $g_{\rm c,eff}^{\rm bulk}$ by the following relation 
		\begin{align}
			&\ab (g_\textrm{c}^{\prime},g_\textrm{s}^{\prime},\theta)  \equiv  \ab (g_{\rm c,eff}^{\rm bulk},1,0) = \frac{1}{g_{\rm c,eff}^{\rm bulk}} -1.
		\end{align}
		It leads to the following definition for $g_{\rm c,eff}^{\rm bulk}$,
		\begin{align}
			\frac{1}{g_{\rm c, eff}^{\rm bulk}} \equiv& \left( \frac{1}{g_\textrm{c}^{\prime}} + \frac{1}{g_\textrm{s}^{\prime}} \right) \left( \cos^2 \theta + g_0^2 \sin^2 \theta \right) -1,
			\label{Eq:gc_eff}
		\end{align}
		which describes the relation between the apparent interaction parameter $g_{\rm c, eff}$ (corresponding to the extracted $g_{\rm c}$ from our experimental observation) and the intrinsic parameters $g_\textrm{c}$, $g_\textrm{s}$, and $\delta v$.
		We remark that the exponent of the power-law conductance does not depend on the number of the barriers, so the definition of  $g_{\rm c,eff}$ is the same for single and multiple barriers in the wire.
				
\begin{figure}[h]
		\includegraphics[width=\linewidth]{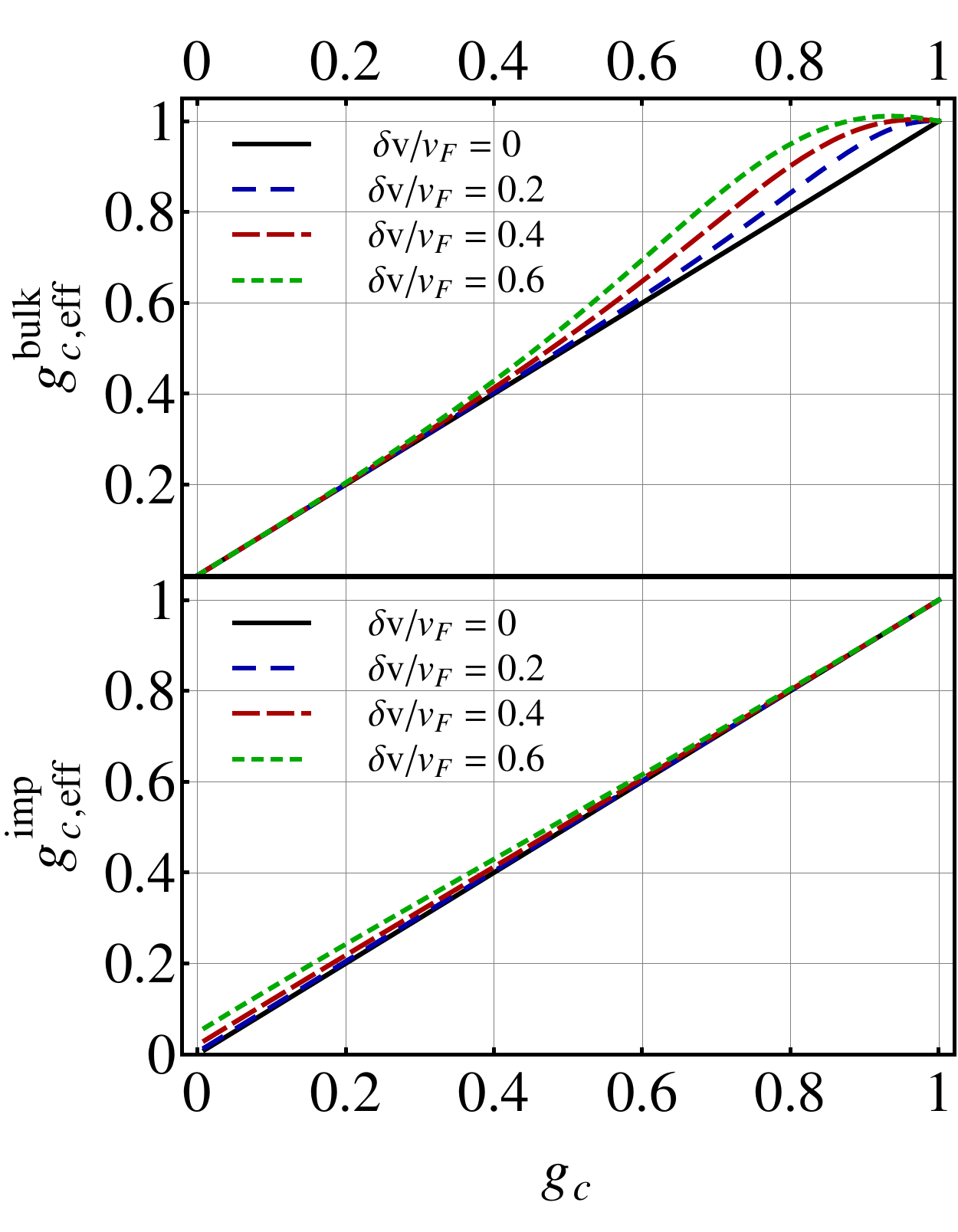}
		\caption{
			Effective interaction parameter $g_{\rm c, eff}$ as a function of the actual interaction parameter $g_\textrm{c}$ for various values of the ratio $\delta v/v_F$. 
			Top: in the case of barriers, $g_{\rm c, eff}^{\rm bulk} $ is defined in Eq.~\eqref{Eq:gc_eff}.
			Bottom: in the case of weak impurities, $g_{\rm c, eff}^{\rm imp}$ defined in Eq.~\eqref{Eq:gc_eff_imp}.
		}
		\label{Fig:gceff}
	\end{figure}

		To visualize the effects of the SOIs on $g_{\rm c, eff}^{\rm bulk}$, we plot it as a function of $g_\textrm{c}$ for several values of $\delta v/v_F$, as displayed in the top panel of Fig.~\ref{Fig:gceff}.
		Note that we intentionally include exaggerated values of $\delta v/v_F \ge 0.2$ in the plot; a more realistic value $\delta v/v_F \lesssim 0.1$ lead to barely visible changes.
		Further, while rather strong SOIs do modify the parameter $g_{\rm c, eff}^{\rm bulk}$, we find two important features relevant to our experiments.
		First, $g_{\rm c,eff}^{\rm bulk}$ increases with an increasing strength of SOIs.
		Therefore, the SOIs cannot lead to a smaller value of the apparent interaction constant $g_{\rm c,eff}^{\rm bulk}$.
		Second, the SOI-induced increase of $g_{\rm c,eff}^{\rm bulk}$ is sizable in the weak- or moderate-interaction regime $(0.5 \le g_\textrm{c} \le 1)$, but becomes negligible for the strong-interaction regime $(g_\textrm{c} \le 0.5)$.
		Thus, these features allow us to neglect the SOIs when extracting the value of $g_\textrm{c}$ in the case of bulk barriers.
		We emphasize that such an approximation is more accurate (becoming {\it almost exact}) in the low-$V_g$ (small $g_\textrm{c}$) regime, which is of our primary interest.

		We now move on to a strong impurity located around a boundary of the wire, which acts as a TLL-Fermi liquid (FL) junction.
		Again, we compute the tunnel current for generic temperatures and bias voltages, as well as the power-law conductance $G_{\rm end} \propto {\rm Max}(k_B T, eV)^{\aed}$ in the high-temperature and high-bias limits.
		In this case, the tunnel current and the power-law conductance are the same as those for a bulk barrier, except that the exponent reads
		\begin{align}
			\aed = & 
			\left( \frac{1}{2g_\textrm{c}^{\prime}} + \frac{1}{2g_\textrm{s}^{\prime}} \right) \left( \cos^2 \theta + g_0^2 \sin^2 \theta \right) -1, 
			\label{Eq:alpha_boundary}
		\end{align}
		from which we can define the same parameter $g_{\rm c,eff}$ as in Eq.~\eqref{Eq:gc_eff}.
		Again, in the presence of several barriers, we have the same exponent $\aed$ and $V \rightarrow \gamma V$.
		Similar to $\ab$, a realistic value of the ratio $\delta v/ v_F$ does not cause a substantial modification of the $\aed$ values, justifying our procedure on the extraction of the $g_{\rm c}$ value using the zero-SOI formula [Eq.~\eqref{Eq:aed_noSOI} in the main text]. 
		
		We now turn to the case of weak impurities, modeled as a backscattering potential.
		In this case, the calculation for a tunnel barrier in Ref.~\cite{Balents1999} is not applicable. 
		We therefore compute the conductance using the method of Refs.~\cite{Kane1992,Giamarchi2003a}.
		Importantly, when there are many impurities of this type (which is true for our relatively long wires compared to the bulk mean free path), the resistance due to the impurities eventually dominates over the contact resistance between the wire and the lead $h/(2e^2)$, leading to a power-law conductance~\cite{Giamarchi2003a}.
		Therefore, we compute the power-law conductance and the corresponding exponent in the high-temperature and high-bias limits. We get
		\begin{subequations} 
			\begin{align}
				&G_{\rm imp} (T,V) \propto {\rm Max}(k_B T, eV)^{\ai},\\
				&\alpha_{\rm imp} = 2 - \cos^2 \theta (g_\textrm{c}^{\prime} + g_\textrm{s}^{\prime} ) - g_0^2 \sin^2 \theta \left( \frac{1}{g_\textrm{c}^{\prime}} + \frac{1}{g_\textrm{s}^{\prime}} \right).
				\label{Eq:alpha_imp} 
			\end{align}
		\end{subequations} 
		Since the conductance is similar to the tunnel barrier case (upon replacing the exponent $\ab/\aed \rightarrow \alpha_{\rm imp}$), the power-law conductance can mimic the scaling behavior observed in our experiment. 
		We therefore take $\alpha \rightarrow \alpha_{\rm imp}$ in Eq.~\eqref{Eq:USC} and treat it as an interpolation formula for the current-voltage curve of a TLL in the presence of weak impurities. 
		In this case, however, $ V$ is the voltage difference across the entire wire instead of a single barrier, so $\gamma = 1$ regardless of the number of impurities. 
		Equation~\eqref{Eq:alpha_imp} allows us to define the effective interaction parameter for the weak-impurity case,
		\begin{align}
			g_{\rm c, eff}^{\rm imp} \equiv& \cos^2 \theta (g_\textrm{c}^{\prime} + g_\textrm{s}^{\prime} ) + g_0^2 \sin^2 \theta \left( \frac{1}{g_\textrm{c}^{\prime}} + \frac{1}{g_\textrm{s}^{\prime}} \right) -1.
			\label{Eq:gc_eff_imp}
		\end{align}
		In the bottom panel of Fig.~\ref{Fig:gceff}, we plot $g_{\rm c, eff}^{\rm imp}$ vs $g_\textrm{c}$.
		We see that the value of $g_{\rm c, eff}^{\rm imp}$ is barely changed, so neither in this case the SOIs lead to substantial effects on the extracted value of the interaction parameter. 
				
		In summary, for all the types of resistance sources we consider here, the effects of SOIs on the extracted value of $g_\textrm{c}$ are negligible.
		Therefore, the experimental values of $g_\textrm{c}$ can be extracted using equations without including the spin-orbit effects, as given in the main text.

	\subsection{Extracting the interaction parameters in various scenarios} 
		In this section we discuss how the theoretically developed results in the previous section are applied to our experimental data, in order to extract the interaction parameters of our quantum wires.
		Since SOIs lead to negligible changes in the parameters $\alpha$ characterizing the universal scaling formula (looking at Fig.~\ref{Fig:gceff}, we find that $g_{\rm c, eff}$ departs from $g_c$ negligibly for anticipated values of the SOI strength), in the following we use the zero-SOI forms of $\alpha$.
		We express $\alpha$ as a function of $g_\textrm{c}$ considering the resistance contributions arising from two tunnel barriers and many weak impurities.
		The former is suggested by the observed value of $1/\gamma \simeq 2$, and the latter is believed to be present since our wires are relatively long on the scale of the bulk mean free path.
		
		We examine the following scenarios: (A) both the two barriers are in the bulk, and (B) they are around the boundaries of the wire (between the TLL and FL). 
		For each of them, we also include the resistance contributions from weak impurities.
		We assume that these resistance sources are independent and their contributions to the total resistance are additive.
		
		In Scenario A, there are weak impurities and two tunnel barriers in a wire.
		Both of them contribute to the total (series) resistance (assuming the contact resistance is much smaller than these two),
		\begin{align}
			R_{\rm a} (T,V) = &\, \frac{1}{G_{\rm bulk} (T,V) }  +  \frac{1}{G_{\rm imp} (T,V) } \nonumber\\ 
			= & \, C_1 \times \frac{h}{e^2} \left[\frac{\Delta_a}{{\rm Max}(k_B T, eV/2)}\right]^{\ab} \nonumber \\
			  & + C_2 \times \frac{h}{2e^2} \left[\frac{\Delta_a}{{\rm Max}(k_B T, eV)}\right]^{\ai}.
		\end{align}
		Here, $\Delta_a$ is the effective bandwidth introduced in the bosonization scheme.
		Assuming that the two dimensionless prefactors ($C_1$ and $C_2$) are of the same order, the relative magnitude of the two resistance contributions is determined by the exponents $\ab$ and $\ai$.
		Because under experimental conditions $\Delta_a$ is much larger than $k_B T$ and $eV$, the term with the larger exponent dominates.
		We therefore examine when the resistance due to the bulk barriers dominates (that is, when $\ab \ge \ai$).
		It leads to the following condition, 
		  \begin{align}
			\ab \gtrsim \ai ~\Leftrightarrow~ g_\textrm{c} \lesssim 1,
		\end{align}
		where the approximation arises from the assumptions of $O(C_{1}) = O( C_{2})$, $g_\textrm{s}=1$, and negligible effects from SOIs.
				
		Therefore, when the tunnel barriers are in the bulk, the contribution from the barriers dominates over the one from weak impurities for any repulsive interaction.
		Consequently, in Scenario A, the impurity-induced resistance is negligible, and we obtain the conductance
		\begin{align}
			G_{\rm a} (T,V)  =&\, \frac{1}{R_{\rm a} (T,V)}  \\
			\approx &\, G_{\rm bulk}(T,V)  \propto  {\rm Max}(k_B T, eV / 2)^{\ab}. \nonumber 
		\end{align}
		We get the universal scaling formula in Eq.~\eqref{Eq:USC}, with $\gamma =1/2$ and $\alpha= \ab$.
		In the main text, we therefore use Eq.~\eqref{Eq:gcTT} to extract the $g_\textrm{c}$ value.
		
		We now turn to Scenario B, in which there are coexisting two tunnel barriers around the boundaries of the wire and many weak impurities. 
		We get
		\begin{align}
			R_{\rm b} (T,V) = & \frac{1}{G_{\rm end} (T,V) }  +  \frac{1}{G_{\rm imp} (T,V) } \\ 
			= & {C_3} \times \frac{h}{e^2} \left[\frac{\Delta_a}{{\rm Max}(k_B T,  eV/2)}\right]^{\aed} \nonumber \\
			  & + {C_2} \times \frac{h}{2e^2} \left[\frac{\Delta_a}{{\rm Max}(k_B T, eV)}\right]^{\ai}. \nonumber
		\end{align}
		The condition for the dominant contribution from the barriers follows as,
		\begin{align}
			\aed  \gtrsim & \ai  ~\Leftrightarrow~ g_\textrm{c} \lesssim \frac{1}{2}.
		\end{align} 
		As a result, there is a transition of the dominant resistance source when varying $g_\textrm{c}$ through the top gate voltage.
		The dominant source changes from the tunnel barriers in the strong-interaction regime ($g_\textrm{c} \le 1/2$) to the weak impurities in the weak-interaction regime ($g_\textrm{c} \ge 1/2$).
		We get
		\begin{align}
			G_{\rm b} (T,V) \propto & \left\{
			\begin{array}{ll}
				{\rm Max}(k_B T, eV/2)^{\aed}, 
				& \textrm{for } g_\textrm{c} \le 1/2, \\
				{\rm Max}(k_B T, eV)^{\ai},
				& \textrm{for }  g_\textrm{c} \ge 1/2,
			\end{array}
			\right.
		\end{align}
		with the exponents $\aed$ and $\ai$ given in Eqs.~\eqref{Eq:alpha_boundary} and \eqref{Eq:alpha_imp}, respectively.
		
		Interestingly, our calculation also suggests that the parameter $\alpha$ should be larger than 0.5 as $g_\textrm{c} \le 1/2$, and smaller than 0.5 as $g_\textrm{c} \ge 1/2$.
		Therefore, in Scenario B we are able to identify the transition of the dominant resistance source based on the observed $\alpha$ values.
		When $\alpha \ge 0.5$, the resistance is due to the boundary tunnel barriers in the regime $g_\textrm{c} \le 1/2$, and therefore $\gamma \simeq 0.5$ in this regime due to the voltage drop shared by the two barriers.
		On the other hand, when $\alpha \le 0.5$, the resistance arises from the impurities, so $\gamma$ becomes unity in this regime.
		
		Consequently, the interpolation formula for Scenario B is given by Eq.~\eqref{Eq:USC}, with the parameters
		 \begin{align}
			(\alpha,\gamma) & =
			\left\{
				\begin{array}{ll}
					(\aed, 1/2), & \mbox{for $\alpha \ge 0.5$ (barriers);} \\
					 (\ai,1), & \mbox{for $\alpha \le 0.5$ (impurities).}
				\end{array}
			\right.
		\end{align}       
		In the main text, we use Eq.~\eqref{Eq:gcTF} to extract the value for $g_\textrm{c}$ from the observed $\alpha$ values.
		Remarkably, upon increasing $V_g$, we observe that $\alpha$ decreases across 0.5 {\it around the same $V_g$ value} at which $\gamma$ changes from $\simeq 0.5$ toward unity.
		The observation is consistent with Scenario B, which provides an explanation of the change in $\gamma$.
		
		Finally, we comment on a third scenario. 
		Namely, one may wonder whether it is possible to have both types of barriers, bulk and boundary. 
		In contrast to our observation, however, this scenario would give conductance with different power laws in the high-bias and high-temperature limits, as discussed in Ref.~\cite{Venkataraman2006}.
		We therefore conclude that this scenario is not relevant to our observations.
		
		In summary, the combined experimental and theoretical results for the considered scenarios indicate that our extracted value of $g_\textrm{c} =$0.16--0.28 is not an artifact of the strong SOIs in InAs wires.
		This conclusion holds regardless whether Scenario A or B is realized.

\renewcommand{\arraystretch}{1.2}
	
	\begin{table*}[t]
		\centering
		\caption{
			Exponent $\alpha$ of the power-law conductance in various TLLs subject to tunnel barriers and many weak impurities (treated as weak potential disorder).
		  The first column lists the system types. 
			The second (fourth) column corresponds to the exponent $\ab$ ($\aed$) for a TLL-TLL (TLL-FL) junction.
			The sixth column lists the exponents $\ai$ corresponding to many weak impurities.
			The references corresponding to the entries are given in the third, fifth, and seventh columns.
			The eighth column lists the allowed ranges for $\ai$, assuming that only one of the sectors is interacting (with the interaction parameters of the other sectors set to unity).
			In the entries, the notation $g$ denotes the interaction parameter in a spinless TLL, while the notation $g_\textrm{c/s}$ denotes the interaction parameter of the charge/spin sectors, respectively, in a spinful TLL (no SOIs).
			For a spinful TLL with the valley degrees of freedom (for example, a carbon nanotube), the notation $g_{\nu P}$ denotes the sectors of the charge/spin degrees of freedom (with $\nu \in \{c,s\}$, respectively), and the symmetric/antisymmetric combination of the valleys (with $P \in \{S,A\}$, respectively).
			The quantities after the approximation symbols ($\approx$) indicate the values of the exponents with $g_\textrm{s}$, $g_\textrm{cA}$, $g_\textrm{sS}$, $g_\textrm{sA}$ set to unity.
			Relevant references are given in the footnotes below the table.
			}
		\begin{tabular}[c]{ l || c | c | c | c | c | c | c}
			TLL type & bulk barrier & Refs. & boundary barrier & Refs. & weak impurities & Refs. & allowed range\footnote{Assuming that $g,~ g_\textrm{c}, ~ g_\textrm{cS} \in [0,1]$.}\\			& $\ab$ &&$\aed$ & & $\ai$ & & for $\ai$\\
			\hhline{=#=|=|=|=|=|=|=}
			&&&&&&\\
			spinless & ${2}{g^{-1}} - 2$ &\cite{Kane1992a,Giamarchi2003a} & ${g^{-1}} - 1$ & \cite{Giamarchi2003a} & $2 -2g$ &\cite{Kane1992,Maslov1995,Giamarchi2003a} & $[0,~ 2]$\\
			&&&&&&\\
			\hline
			&&&&&&\\
			spinful & $g_\textrm{c}^{-1} + g_\textrm{s}^{-1} - 2 $&\cite{Kane1992} &  
			$\frac{1}{2} (g_\textrm{c}^{-1} + g_\textrm{s}^{-1}) - 1 $&\cite{Matveev1993a}
			&$2 - g_\textrm{c} - g_\textrm{s}$ & \cite{Kane1992,Maslov1995}\footnote{See also the calculation in the presence of the multibands or multiple-channels\cite{Sandler1997}.} & $ [0, 1]$\\
			& $\approx g_\textrm{c}^{-1} -1 $ &&  
			$\approx \frac{1}{2} (g_\textrm{c}^{-1} -1)$&&
			$\approx 1 - g_\textrm{c}$&\\		
			&&&&&&\\
			\hline
		 	&&&&&&\\
			spinful with & 
			$ \frac{1}{2}  g^{\rm inv}_{\rm sum} - 2  $& \cite{Yao1999}\footnote{See also the calculation for multi-wall nanotubes or ropes of single-wall nanotubes\cite{Egger1999}.}
			&
			$ \frac{1}{4} g^{\rm inv}_{\rm sum} - 1$&\cite{Kane1997,Balents1999,Bockrath1999} 
			&$2- \frac{1}{2}  g_{\rm sum}$ &\cite{Egger1997,Kane1997}\footnotemark[3]
			& $[0,~1/2]$\\	
			two valleys\footnote{For this entry, we define $g_{\rm sum}=g_\textrm{cS} + g_\textrm{cA} + g_\textrm{sS} + g_\textrm{sA}$ and $g^{\rm inv}_{\rm sum}=g_\textrm{cS}^{-1} + g_\textrm{cA}^{-1} + g_\textrm{sS}^{-1} + g_\textrm{sA}^{-1}$.}
			& 
			$   \approx \frac{1}{2}  ( g_\textrm{cS}^{-1} - 1 ) $
			&&
			$\approx \frac{1}{4}  ( g_\textrm{cS}^{-1} - 1 ) $&
			&$\approx \frac{1}{2}  ( 1- g_\textrm{cS} ) $&
			& \\	
			&&&&&&
		\end{tabular}
		\label{Tab:tunnel}
		\end{table*}

	\subsection{Summary of the power-law conductance in various TLLs}
	
		In this section we give a summary of the parameter $\alpha$ for various TLL systems.
		It allows us to conduct a survey on the interaction parameters in various one-dimensional systems in literature listed in Table~\ref{Tab:gc} in the main text.
		To be specific, the summary includes a spinless TLL, a spinful TLL (without SOIs), and a spinful TLL with valley degrees of freedom (that is, a carbon nanotube).

In Table~\ref{Tab:tunnel}, we list the exponent $\ab/\aed/\ai$ for various TLLs subject to tunnel barriers and many weak impurities.
		We remark that the exponents $\ab/\aed$ are the same for single and multiple barriers in the wire.
		The first column gives the system type.
		The second column corresponds to the scenario in which the tunnel barriers (isolated, strong impurities) located in the bulk of a wire (that is, TLL-TLL junctions), with the references given in the third column.
		The fourth column corresponds to the tunnel barriers are located around the boundaries of the wire (that is, TLL-FL junctions), with the corresponding references in the fifth column.
		The sixth column lists the exponent $\ai$ for various TLLs subject to many weak impurities, with the references in the seventh column.
		In contrast to $\ab/\aed$ in the tunneling regime, the value of $\ai$ is bounded.
		In the table we give the allowed ranges, assuming that the electron-electron interactions only act on one sector.
		For example, for a spinful TLL, we assume that the spin sector is noninteracting, which fixes $g_\textrm{s}=1$. 
		For repulsive interactions, the interaction parameter of the charge sector is in the range $g_\textrm{c} \in [0,1]$, leading to a bound $\alpha_{\rm imp} \in [0,1]$. 
		We remark that, since the value $\ai \le 1$ is inconsistent with the observed $\alpha$ value in the low-density (low-$V_g$) regime in our experiment, we are able to attribute the observed power-law behavior in that regime to the tunnel barriers and rule out weak potential disorder.

%%%%%%%%%%%%%%%%%%%%%%%%%%%%%%%%%%%%%%%%%%%

%\bibliography{library,add}

\begin{thebibliography}{79}%
\makeatletter
\providecommand \@ifxundefined [1]{%
 \@ifx{#1\undefined}
}%
\providecommand \@ifnum [1]{%
 \ifnum #1\expandafter \@firstoftwo
 \else \expandafter \@secondoftwo
 \fi
}%
\providecommand \@ifx [1]{%
 \ifx #1\expandafter \@firstoftwo
 \else \expandafter \@secondoftwo
 \fi
}%
\providecommand \natexlab [1]{#1}%
\providecommand \enquote  [1]{``#1''}%
\providecommand \bibnamefont  [1]{#1}%
\providecommand \bibfnamefont [1]{#1}%
\providecommand \citenamefont [1]{#1}%
\providecommand \href@noop [0]{\@secondoftwo}%
\providecommand \href [0]{\begingroup \@sanitize@url \@href}%
\providecommand \@href[1]{\@@startlink{#1}\@@href}%
\providecommand \@@href[1]{\endgroup#1\@@endlink}%
\providecommand \@sanitize@url [0]{\catcode `\\12\catcode `\$12\catcode
  `\&12\catcode `\#12\catcode `\^12\catcode `\_12\catcode `\%12\relax}%
\providecommand \@@startlink[1]{}%
\providecommand \@@endlink[0]{}%
\providecommand \url  [0]{\begingroup\@sanitize@url \@url }%
\providecommand \@url [1]{\endgroup\@href {#1}{\urlprefix }}%
\providecommand \urlprefix  [0]{URL }%
\providecommand \Eprint [0]{\href }%
\providecommand \doibase [0]{http://dx.doi.org/}%
\providecommand \selectlanguage [0]{\@gobble}%
\providecommand \bibinfo  [0]{\@secondoftwo}%
\providecommand \bibfield  [0]{\@secondoftwo}%
\providecommand \translation [1]{[#1]}%
\providecommand \BibitemOpen [0]{}%
\providecommand \bibitemStop [0]{}%
\providecommand \bibitemNoStop [0]{.\EOS\space}%
\providecommand \EOS [0]{\spacefactor3000\relax}%
\providecommand \BibitemShut  [1]{\csname bibitem#1\endcsname}%
\let\auto@bib@innerbib\@empty
%</preamble>
\bibitem [{\citenamefont {Tomonaga}(1950)}]{Tomonaga1950}%
  \BibitemOpen
  \bibfield  {author} {\bibinfo {author} {\bibfnamefont {S.-i.}\ \bibnamefont
  {Tomonaga}},\ }\href {\doibase 10.1143/ptp/5.4.544} {\bibfield  {journal}
  {\bibinfo  {journal} {Prog. Theor. Phys.}\ }\textbf {\bibinfo {volume} {5}},\
  \bibinfo {pages} {544} (\bibinfo {year} {1950})}\BibitemShut {NoStop}%
\bibitem [{\citenamefont {Luttinger}(1963)}]{Luttinger1963}%
  \BibitemOpen
  \bibfield  {author} {\bibinfo {author} {\bibfnamefont {J.~M.}\ \bibnamefont
  {Luttinger}},\ }\href {\doibase 10.1063/1.1704046} {\bibfield  {journal}
  {\bibinfo  {journal} {J. Math. Phys.}\ }\textbf {\bibinfo {volume} {4}},\
  \bibinfo {pages} {1154} (\bibinfo {year} {1963})}\BibitemShut {NoStop}%
\bibitem [{\citenamefont {Matveev}(2004)}]{Matveev2004}%
  \BibitemOpen
  \bibfield  {author} {\bibinfo {author} {\bibfnamefont {K.~A.}\ \bibnamefont
  {Matveev}},\ }\href {\doibase 10.1103/PhysRevB.70.245319} {\bibfield
  {journal} {\bibinfo  {journal} {Phys. Rev. B}\ }\textbf {\bibinfo {volume}
  {70}},\ \bibinfo {pages} {245319} (\bibinfo {year} {2004})},\ \Eprint
  {http://arxiv.org/abs/0405542} {arXiv:0405542 [cond-mat]} \BibitemShut
  {NoStop}%
\bibitem [{\citenamefont {Furusaki}\ and\ \citenamefont
  {Nagaosa}(1993)}]{Furusaki1993a}%
  \BibitemOpen
  \bibfield  {author} {\bibinfo {author} {\bibfnamefont {A.}~\bibnamefont
  {Furusaki}}\ and\ \bibinfo {author} {\bibfnamefont {N.}~\bibnamefont
  {Nagaosa}},\ }\href {\doibase 10.1103/PhysRevB.47.4631} {\bibfield  {journal}
  {\bibinfo  {journal} {Phys. Rev. B}\ }\textbf {\bibinfo {volume} {47}},\
  \bibinfo {pages} {4631} (\bibinfo {year} {1993})}\BibitemShut {NoStop}%
\bibitem [{\citenamefont {Datta}\ and\ \citenamefont {Das}(1990)}]{Datta1990}%
  \BibitemOpen
  \bibfield  {author} {\bibinfo {author} {\bibfnamefont {S.}~\bibnamefont
  {Datta}}\ and\ \bibinfo {author} {\bibfnamefont {B.}~\bibnamefont {Das}},\
  }\href {\doibase 10.1063/1.102730} {\bibfield  {journal} {\bibinfo  {journal}
  {Appl. Phys. Lett.}\ }\textbf {\bibinfo {volume} {56}},\ \bibinfo {pages}
  {665} (\bibinfo {year} {1990})}\BibitemShut {NoStop}%
\bibitem [{\citenamefont {Fabian}\ \emph {et~al.}(2007)\citenamefont {Fabian},
  \citenamefont {Matos-Abiague}, \citenamefont {Ertler}, \citenamefont
  {Stano},\ and\ \citenamefont {{\v{Z}}uti{\'{c}}}}]{Fabian2007}%
  \BibitemOpen
  \bibfield  {author} {\bibinfo {author} {\bibfnamefont {J.}~\bibnamefont
  {Fabian}}, \bibinfo {author} {\bibfnamefont {A.}~\bibnamefont
  {Matos-Abiague}}, \bibinfo {author} {\bibfnamefont {C.}~\bibnamefont
  {Ertler}}, \bibinfo {author} {\bibfnamefont {P.}~\bibnamefont {Stano}}, \
  and\ \bibinfo {author} {\bibfnamefont {I.}~\bibnamefont
  {{\v{Z}}uti{\'{c}}}},\ }\href {\doibase 10.2478/v10155-010-0086-8} {\bibfield
   {journal} {\bibinfo  {journal} {Acta Phys. Slovaca. Rev. Tutorials}\
  }\textbf {\bibinfo {volume} {57}},\ \bibinfo {pages} {565} (\bibinfo {year}
  {2007})}\BibitemShut {NoStop}%
\bibitem [{\citenamefont {Lutchyn}\ \emph {et~al.}(2010)\citenamefont
  {Lutchyn}, \citenamefont {Sau},\ and\ \citenamefont {{Das
  Sarma}}}]{Lutchyn2010}%
  \BibitemOpen
  \bibfield  {author} {\bibinfo {author} {\bibfnamefont {R.~M.}\ \bibnamefont
  {Lutchyn}}, \bibinfo {author} {\bibfnamefont {J.~D.}\ \bibnamefont {Sau}}, \
  and\ \bibinfo {author} {\bibfnamefont {S.}~\bibnamefont {{Das Sarma}}},\
  }\href {\doibase 10.1103/PhysRevLett.105.077001} {\bibfield  {journal}
  {\bibinfo  {journal} {Phys. Rev. Lett.}\ }\textbf {\bibinfo {volume} {105}},\
  \bibinfo {pages} {077001} (\bibinfo {year} {2010})},\ \Eprint
  {http://arxiv.org/abs/1002.4033} {arXiv:1002.4033} \BibitemShut {NoStop}%
\bibitem [{\citenamefont {Oreg}\ \emph {et~al.}(2010)\citenamefont {Oreg},
  \citenamefont {Refael},\ and\ \citenamefont {von Oppen}}]{Oreg2010}%
  \BibitemOpen
  \bibfield  {author} {\bibinfo {author} {\bibfnamefont {Y.}~\bibnamefont
  {Oreg}}, \bibinfo {author} {\bibfnamefont {G.}~\bibnamefont {Refael}}, \ and\
  \bibinfo {author} {\bibfnamefont {F.}~\bibnamefont {von Oppen}},\ }\href
  {\doibase 10.1103/PhysRevLett.105.177002} {\bibfield  {journal} {\bibinfo
  {journal} {Phys. Rev. Lett.}\ }\textbf {\bibinfo {volume} {105}},\ \bibinfo
  {pages} {177002} (\bibinfo {year} {2010})},\ \Eprint
  {http://arxiv.org/abs/1003.1145} {arXiv:1003.1145} \BibitemShut {NoStop}%
\bibitem [{\citenamefont {Klinovaja}\ \emph {et~al.}(2012)\citenamefont
  {Klinovaja}, \citenamefont {Stano},\ and\ \citenamefont
  {Loss}}]{Klinovaja2012}%
  \BibitemOpen
  \bibfield  {author} {\bibinfo {author} {\bibfnamefont {J.}~\bibnamefont
  {Klinovaja}}, \bibinfo {author} {\bibfnamefont {P.}~\bibnamefont {Stano}}, \
  and\ \bibinfo {author} {\bibfnamefont {D.}~\bibnamefont {Loss}},\ }\href
  {\doibase 10.1103/PhysRevLett.109.236801} {\bibfield  {journal} {\bibinfo
  {journal} {Phys. Rev. Lett.}\ }\textbf {\bibinfo {volume} {109}},\ \bibinfo
  {pages} {236801} (\bibinfo {year} {2012})},\ \Eprint
  {http://arxiv.org/abs/1207.7322} {arXiv:1207.7322} \BibitemShut {NoStop}%
\bibitem [{\citenamefont {Alicea}(2010)}]{Alicea2010}%
  \BibitemOpen
  \bibfield  {author} {\bibinfo {author} {\bibfnamefont {J.}~\bibnamefont
  {Alicea}},\ }\href {\doibase 10.1103/PhysRevB.81.125318} {\bibfield
  {journal} {\bibinfo  {journal} {Phys. Rev. B}\ }\textbf {\bibinfo {volume}
  {81}},\ \bibinfo {pages} {125318} (\bibinfo {year} {2010})},\ \Eprint
  {http://arxiv.org/abs/0912.2115} {arXiv:0912.2115} \BibitemShut {NoStop}%
\bibitem [{\citenamefont {Klinovaja}\ and\ \citenamefont
  {Loss}(2014)}]{Klinovaja2014}%
  \BibitemOpen
  \bibfield  {author} {\bibinfo {author} {\bibfnamefont {J.}~\bibnamefont
  {Klinovaja}}\ and\ \bibinfo {author} {\bibfnamefont {D.}~\bibnamefont
  {Loss}},\ }\href {\doibase 10.1103/PhysRevB.90.045118} {\bibfield  {journal}
  {\bibinfo  {journal} {Phys. Rev. B}\ }\textbf {\bibinfo {volume} {90}},\
  \bibinfo {pages} {045118} (\bibinfo {year} {2014})},\ \Eprint
  {http://arxiv.org/abs/arXiv:1312.1998v1} {arXiv:arXiv:1312.1998v1}
  \BibitemShut {NoStop}%
\bibitem [{\citenamefont {Thakurathi}\ \emph {et~al.}(2018)\citenamefont
  {Thakurathi}, \citenamefont {Simon}, \citenamefont {Mandal}, \citenamefont
  {Klinovaja},\ and\ \citenamefont {Loss}}]{Thakurathi2018}%
  \BibitemOpen
  \bibfield  {author} {\bibinfo {author} {\bibfnamefont {M.}~\bibnamefont
  {Thakurathi}}, \bibinfo {author} {\bibfnamefont {P.}~\bibnamefont {Simon}},
  \bibinfo {author} {\bibfnamefont {I.}~\bibnamefont {Mandal}}, \bibinfo
  {author} {\bibfnamefont {J.}~\bibnamefont {Klinovaja}}, \ and\ \bibinfo
  {author} {\bibfnamefont {D.}~\bibnamefont {Loss}},\ }\href {\doibase
  10.1103/PhysRevB.97.045415} {\bibfield  {journal} {\bibinfo  {journal} {Phys.
  Rev. B}\ }\textbf {\bibinfo {volume} {97}},\ \bibinfo {pages} {045415}
  (\bibinfo {year} {2018})},\ \Eprint {http://arxiv.org/abs/1711.04682}
  {arXiv:1711.04682} \BibitemShut {NoStop}%
\bibitem [{\citenamefont {Fendley}(2012)}]{Fendley2012}%
  \BibitemOpen
  \bibfield  {author} {\bibinfo {author} {\bibfnamefont {P.}~\bibnamefont
  {Fendley}},\ }\href {\doibase 10.1088/1742-5468/2012/11/P11020} {\bibfield
  {journal} {\bibinfo  {journal} {J. Stat. Mech. Theory Exp.}\ }\textbf
  {\bibinfo {volume} {2012}},\ \bibinfo {pages} {P11020} (\bibinfo {year}
  {2012})},\ \Eprint {http://arxiv.org/abs/1209.0472} {arXiv:1209.0472}
  \BibitemShut {NoStop}%
\bibitem [{\citenamefont {Hutter}\ and\ \citenamefont
  {Loss}(2016)}]{Hutter2016}%
  \BibitemOpen
  \bibfield  {author} {\bibinfo {author} {\bibfnamefont {A.}~\bibnamefont
  {Hutter}}\ and\ \bibinfo {author} {\bibfnamefont {D.}~\bibnamefont {Loss}},\
  }\href {\doibase 10.1103/PhysRevB.93.125105} {\bibfield  {journal} {\bibinfo
  {journal} {Phys. Rev. B}\ }\textbf {\bibinfo {volume} {93}},\ \bibinfo
  {pages} {125105} (\bibinfo {year} {2016})},\ \Eprint
  {http://arxiv.org/abs/1511.02704} {arXiv:1511.02704} \BibitemShut {NoStop}%
\bibitem [{\citenamefont {Baba}\ \emph {et~al.}(2018)\citenamefont {Baba},
  \citenamefont {J{\"{u}}nger}, \citenamefont {Matsuo}, \citenamefont
  {Baumgartner}, \citenamefont {Sato}, \citenamefont {Kamata}, \citenamefont
  {Li}, \citenamefont {Jeppesen}, \citenamefont {Samuelson}, \citenamefont
  {Xu}, \citenamefont {Sch{\"{o}}nenberger},\ and\ \citenamefont
  {Tarucha}}]{Baba2018}%
  \BibitemOpen
  \bibfield  {author} {\bibinfo {author} {\bibfnamefont {S.}~\bibnamefont
  {Baba}}, \bibinfo {author} {\bibfnamefont {C.}~\bibnamefont {J{\"{u}}nger}},
  \bibinfo {author} {\bibfnamefont {S.}~\bibnamefont {Matsuo}}, \bibinfo
  {author} {\bibfnamefont {A.}~\bibnamefont {Baumgartner}}, \bibinfo {author}
  {\bibfnamefont {Y.}~\bibnamefont {Sato}}, \bibinfo {author} {\bibfnamefont
  {H.}~\bibnamefont {Kamata}}, \bibinfo {author} {\bibfnamefont
  {K.}~\bibnamefont {Li}}, \bibinfo {author} {\bibfnamefont {S.}~\bibnamefont
  {Jeppesen}}, \bibinfo {author} {\bibfnamefont {L.}~\bibnamefont {Samuelson}},
  \bibinfo {author} {\bibfnamefont {H.~Q.}\ \bibnamefont {Xu}}, \bibinfo
  {author} {\bibfnamefont {C.}~\bibnamefont {Sch{\"{o}}nenberger}}, \ and\
  \bibinfo {author} {\bibfnamefont {S.}~\bibnamefont {Tarucha}},\ }\href
  {\doibase 10.1088/1367-2630/aac74e} {\bibfield  {journal} {\bibinfo
  {journal} {New J. Phys.}\ }\textbf {\bibinfo {volume} {20}},\ \bibinfo
  {pages} {063021} (\bibinfo {year} {2018})},\ \Eprint
  {http://arxiv.org/abs/1802.08059} {arXiv:1802.08059} \BibitemShut {NoStop}%
\bibitem [{\citenamefont {Ueda}\ \emph {et~al.}(2018)\citenamefont {Ueda},
  \citenamefont {Matsuo}, \citenamefont {Kamata}, \citenamefont {Baba},
  \citenamefont {Sato}, \citenamefont {Takeshige}, \citenamefont {Li},
  \citenamefont {Jeppesen}, \citenamefont {Samuelson}, \citenamefont {Xu},\
  and\ \citenamefont {Tarucha}}]{Ueda2018}%
  \BibitemOpen
  \bibfield  {author} {\bibinfo {author} {\bibfnamefont {K.}~\bibnamefont
  {Ueda}}, \bibinfo {author} {\bibfnamefont {S.}~\bibnamefont {Matsuo}},
  \bibinfo {author} {\bibfnamefont {H.}~\bibnamefont {Kamata}}, \bibinfo
  {author} {\bibfnamefont {S.}~\bibnamefont {Baba}}, \bibinfo {author}
  {\bibfnamefont {Y.}~\bibnamefont {Sato}}, \bibinfo {author} {\bibfnamefont
  {Y.}~\bibnamefont {Takeshige}}, \bibinfo {author} {\bibfnamefont
  {K.}~\bibnamefont {Li}}, \bibinfo {author} {\bibfnamefont {S.}~\bibnamefont
  {Jeppesen}}, \bibinfo {author} {\bibfnamefont {L.}~\bibnamefont {Samuelson}},
  \bibinfo {author} {\bibfnamefont {H.}~\bibnamefont {Xu}}, \ and\ \bibinfo
  {author} {\bibfnamefont {S.}~\bibnamefont {Tarucha}},\ }\href
  {http://arxiv.org/abs/1810.04832} {\ ,\ \bibinfo {pages} {1} (\bibinfo {year}
  {2018})},\ \Eprint {http://arxiv.org/abs/1810.04832} {arXiv:1810.04832}
  \BibitemShut {NoStop}%
\bibitem [{\citenamefont {Chang}\ \emph {et~al.}(2015)\citenamefont {Chang},
  \citenamefont {Albrecht}, \citenamefont {Jespersen}, \citenamefont
  {Kuemmeth}, \citenamefont {Krogstrup}, \citenamefont {Nyg{\aa}rd},\ and\
  \citenamefont {Marcus}}]{Chang2015}%
  \BibitemOpen
  \bibfield  {author} {\bibinfo {author} {\bibfnamefont {W.}~\bibnamefont
  {Chang}}, \bibinfo {author} {\bibfnamefont {S.~M.}\ \bibnamefont {Albrecht}},
  \bibinfo {author} {\bibfnamefont {T.~S.}\ \bibnamefont {Jespersen}}, \bibinfo
  {author} {\bibfnamefont {F.}~\bibnamefont {Kuemmeth}}, \bibinfo {author}
  {\bibfnamefont {P.}~\bibnamefont {Krogstrup}}, \bibinfo {author}
  {\bibfnamefont {J.}~\bibnamefont {Nyg{\aa}rd}}, \ and\ \bibinfo {author}
  {\bibfnamefont {C.~M.}\ \bibnamefont {Marcus}},\ }\href {\doibase
  10.1038/nnano.2014.306} {\bibfield  {journal} {\bibinfo  {journal} {Nat.
  Nanotechnol.}\ }\textbf {\bibinfo {volume} {10}},\ \bibinfo {pages} {232}
  (\bibinfo {year} {2015})},\ \Eprint {http://arxiv.org/abs/1411.6255}
  {arXiv:1411.6255} \BibitemShut {NoStop}%
\bibitem [{\citenamefont {Krogstrup}\ \emph {et~al.}(2015)\citenamefont
  {Krogstrup}, \citenamefont {Ziino}, \citenamefont {Chang}, \citenamefont
  {Albrecht}, \citenamefont {Madsen}, \citenamefont {Johnson}, \citenamefont
  {Nyg{\aa}rd}, \citenamefont {Marcus},\ and\ \citenamefont
  {Jespersen}}]{Krogstrup2015}%
  \BibitemOpen
  \bibfield  {author} {\bibinfo {author} {\bibfnamefont {P.}~\bibnamefont
  {Krogstrup}}, \bibinfo {author} {\bibfnamefont {N.~L.~B.}\ \bibnamefont
  {Ziino}}, \bibinfo {author} {\bibfnamefont {W.}~\bibnamefont {Chang}},
  \bibinfo {author} {\bibfnamefont {S.~M.}\ \bibnamefont {Albrecht}}, \bibinfo
  {author} {\bibfnamefont {M.~H.}\ \bibnamefont {Madsen}}, \bibinfo {author}
  {\bibfnamefont {E.}~\bibnamefont {Johnson}}, \bibinfo {author} {\bibfnamefont
  {J.}~\bibnamefont {Nyg{\aa}rd}}, \bibinfo {author} {\bibfnamefont {C.~M.}\
  \bibnamefont {Marcus}}, \ and\ \bibinfo {author} {\bibfnamefont {T.~S.}\
  \bibnamefont {Jespersen}},\ }\href {\doibase 10.1038/nmat4176} {\bibfield
  {journal} {\bibinfo  {journal} {Nat. Mater.}\ }\textbf {\bibinfo {volume}
  {14}},\ \bibinfo {pages} {400} (\bibinfo {year} {2015})},\ \Eprint
  {http://arxiv.org/abs/1411.6254} {arXiv:1411.6254 [cond-mat.mtrl-sci]}
  \BibitemShut {NoStop}%
\bibitem [{\citenamefont {Kjaergaard}\ \emph {et~al.}(2016)\citenamefont
  {Kjaergaard}, \citenamefont {Nichele}, \citenamefont {Suominen},
  \citenamefont {Nowak}, \citenamefont {Wimmer}, \citenamefont {Akhmerov},
  \citenamefont {Folk}, \citenamefont {Flensberg}, \citenamefont {Shabani},
  \citenamefont {Palmstr{\o}m},\ and\ \citenamefont {Marcus}}]{Kjaergaard2016}%
  \BibitemOpen
  \bibfield  {author} {\bibinfo {author} {\bibfnamefont {M.}~\bibnamefont
  {Kjaergaard}}, \bibinfo {author} {\bibfnamefont {F.}~\bibnamefont {Nichele}},
  \bibinfo {author} {\bibfnamefont {H.~J.}\ \bibnamefont {Suominen}}, \bibinfo
  {author} {\bibfnamefont {M.~P.}\ \bibnamefont {Nowak}}, \bibinfo {author}
  {\bibfnamefont {M.}~\bibnamefont {Wimmer}}, \bibinfo {author} {\bibfnamefont
  {A.~R.}\ \bibnamefont {Akhmerov}}, \bibinfo {author} {\bibfnamefont {J.~A.}\
  \bibnamefont {Folk}}, \bibinfo {author} {\bibfnamefont {K.}~\bibnamefont
  {Flensberg}}, \bibinfo {author} {\bibfnamefont {J.}~\bibnamefont {Shabani}},
  \bibinfo {author} {\bibfnamefont {C.~J.}\ \bibnamefont {Palmstr{\o}m}}, \
  and\ \bibinfo {author} {\bibfnamefont {C.~M.}\ \bibnamefont {Marcus}},\
  }\href {\doibase 10.1038/ncomms12841} {\bibfield  {journal} {\bibinfo
  {journal} {Nat. Commun.}\ }\textbf {\bibinfo {volume} {7}},\ \bibinfo {pages}
  {12841} (\bibinfo {year} {2016})},\ \Eprint {http://arxiv.org/abs/1603.01852}
  {arXiv:1603.01852} \BibitemShut {NoStop}%
\bibitem [{\citenamefont {Moroz}\ \emph
  {et~al.}(2000{\natexlab{a}})\citenamefont {Moroz}, \citenamefont {Samokhin},\
  and\ \citenamefont {Barnes}}]{Moroz2000}%
  \BibitemOpen
  \bibfield  {author} {\bibinfo {author} {\bibfnamefont {A.~V.}\ \bibnamefont
  {Moroz}}, \bibinfo {author} {\bibfnamefont {K.~V.}\ \bibnamefont {Samokhin}},
  \ and\ \bibinfo {author} {\bibfnamefont {C.~H.~W.}\ \bibnamefont {Barnes}},\
  }\href {\doibase 10.1103/PhysRevLett.84.4164} {\bibfield  {journal} {\bibinfo
   {journal} {Phys. Rev. Lett.}\ }\textbf {\bibinfo {volume} {84}},\ \bibinfo
  {pages} {4164} (\bibinfo {year} {2000}{\natexlab{a}})},\ \Eprint
  {http://arxiv.org/abs/0003219} {arXiv:0003219 [cond-mat]} \BibitemShut
  {NoStop}%
\bibitem [{\citenamefont {Moroz}\ \emph
  {et~al.}(2000{\natexlab{b}})\citenamefont {Moroz}, \citenamefont {Samokhin},\
  and\ \citenamefont {Barnes}}]{Moroz2000a}%
  \BibitemOpen
  \bibfield  {author} {\bibinfo {author} {\bibfnamefont {A.~V.}\ \bibnamefont
  {Moroz}}, \bibinfo {author} {\bibfnamefont {K.~V.}\ \bibnamefont {Samokhin}},
  \ and\ \bibinfo {author} {\bibfnamefont {C.~H.~W.}\ \bibnamefont {Barnes}},\
  }\href {\doibase 10.1103/PhysRevB.62.16900} {\bibfield  {journal} {\bibinfo
  {journal} {Phys. Rev. B}\ }\textbf {\bibinfo {volume} {62}},\ \bibinfo
  {pages} {16900} (\bibinfo {year} {2000}{\natexlab{b}})}\BibitemShut {NoStop}%
\bibitem [{\citenamefont {H\"{a}usler}(2001)}]{Hausler2001}%
  \BibitemOpen
  \bibfield  {author} {\bibinfo {author} {\bibfnamefont {W.}~\bibnamefont
  {H\"{a}usler}},\ }\href {\doibase 10.1103/PhysRevB.63.121310} {\bibfield
  {journal} {\bibinfo  {journal} {Phys. Rev. B}\ }\textbf {\bibinfo {volume}
  {63}},\ \bibinfo {pages} {121310} (\bibinfo {year} {2001})},\ \Eprint
  {http://arxiv.org/abs/0306359} {arXiv:0306359 [cond-mat]} \BibitemShut
  {NoStop}%
\bibitem [{\citenamefont {Martino}\ and\ \citenamefont
  {Egger}(2001)}]{DeMartino2001}%
  \BibitemOpen
  \bibfield  {author} {\bibinfo {author} {\bibfnamefont {A.~D.}\ \bibnamefont
  {Martino}}\ and\ \bibinfo {author} {\bibfnamefont {R.}~\bibnamefont
  {Egger}},\ }\href {\doibase 10.1209/epl/i2001-00558-3} {\bibfield  {journal}
  {\bibinfo  {journal} {Europhys. Lett.}\ }\textbf {\bibinfo {volume} {56}},\
  \bibinfo {pages} {570} (\bibinfo {year} {2001})},\ \Eprint
  {http://arxiv.org/abs/0109161} {arXiv:0109161 [cond-mat]} \BibitemShut
  {NoStop}%
\bibitem [{\citenamefont {Governale}\ and\ \citenamefont
  {Z\"{u}licke}(2002)}]{Governale2002}%
  \BibitemOpen
  \bibfield  {author} {\bibinfo {author} {\bibfnamefont {M.}~\bibnamefont
  {Governale}}\ and\ \bibinfo {author} {\bibfnamefont {U.}~\bibnamefont
  {Z\"{u}licke}},\ }\href {\doibase 10.1103/PhysRevB.66.073311} {\bibfield
  {journal} {\bibinfo  {journal} {Phys. Rev. B}\ }\textbf {\bibinfo {volume}
  {66}},\ \bibinfo {pages} {073311} (\bibinfo {year} {2002})}\BibitemShut
  {NoStop}%
\bibitem [{\citenamefont {Iucci}(2003)}]{Iucci2003}%
  \BibitemOpen
  \bibfield  {author} {\bibinfo {author} {\bibfnamefont {A.}~\bibnamefont
  {Iucci}},\ }\href {\doibase 10.1103/PhysRevB.68.075107} {\bibfield  {journal}
  {\bibinfo  {journal} {Phys. Rev. B}\ }\textbf {\bibinfo {volume} {68}},\
  \bibinfo {pages} {075107} (\bibinfo {year} {2003})}\BibitemShut {NoStop}%
\bibitem [{\citenamefont {Governale}\ and\ \citenamefont
  {Z\"{u}licke}(2004)}]{Governale2004}%
  \BibitemOpen
  \bibfield  {author} {\bibinfo {author} {\bibfnamefont {M.}~\bibnamefont
  {Governale}}\ and\ \bibinfo {author} {\bibfnamefont {U.}~\bibnamefont
  {Z\"{u}licke}},\ }\href {\doibase 10.1016/j.ssc.2004.05.047} {\bibfield
  {journal} {\bibinfo  {journal} {Solid State Commun.}\ }\textbf {\bibinfo
  {volume} {131}},\ \bibinfo {pages} {581} (\bibinfo {year} {2004})},\ \Eprint
  {http://arxiv.org/abs/0407036} {arXiv:0407036 [cond-mat]} \BibitemShut
  {NoStop}%
\bibitem [{\citenamefont {H\"{a}usler}(2004)}]{Hausler2004}%
  \BibitemOpen
  \bibfield  {author} {\bibinfo {author} {\bibfnamefont {W.}~\bibnamefont
  {H\"{a}usler}},\ }\href {\doibase 10.1103/PhysRevB.70.115313} {\bibfield
  {journal} {\bibinfo  {journal} {Phys. Rev. B}\ }\textbf {\bibinfo {volume}
  {70}},\ \bibinfo {pages} {115313} (\bibinfo {year} {2004})}\BibitemShut
  {NoStop}%
\bibitem [{\citenamefont {Sun}\ \emph {et~al.}(2007)\citenamefont {Sun},
  \citenamefont {Gangadharaiah},\ and\ \citenamefont {Starykh}}]{Sun2007}%
  \BibitemOpen
  \bibfield  {author} {\bibinfo {author} {\bibfnamefont {J.}~\bibnamefont
  {Sun}}, \bibinfo {author} {\bibfnamefont {S.}~\bibnamefont {Gangadharaiah}},
  \ and\ \bibinfo {author} {\bibfnamefont {O.~A.}\ \bibnamefont {Starykh}},\
  }\href {\doibase 10.1103/PhysRevLett.98.126408} {\bibfield  {journal}
  {\bibinfo  {journal} {Phys. Rev. Lett.}\ }\textbf {\bibinfo {volume} {98}},\
  \bibinfo {pages} {126408} (\bibinfo {year} {2007})}\BibitemShut {NoStop}%
\bibitem [{\citenamefont {Schulz}\ \emph {et~al.}(2009)\citenamefont {Schulz},
  \citenamefont {{De Martino}}, \citenamefont {Ingenhoven},\ and\ \citenamefont
  {Egger}}]{Schulz2009}%
  \BibitemOpen
  \bibfield  {author} {\bibinfo {author} {\bibfnamefont {A.}~\bibnamefont
  {Schulz}}, \bibinfo {author} {\bibfnamefont {A.}~\bibnamefont {{De
  Martino}}}, \bibinfo {author} {\bibfnamefont {P.}~\bibnamefont {Ingenhoven}},
  \ and\ \bibinfo {author} {\bibfnamefont {R.}~\bibnamefont {Egger}},\ }\href
  {\doibase 10.1103/PhysRevB.79.205432} {\bibfield  {journal} {\bibinfo
  {journal} {Phys. Rev. B}\ }\textbf {\bibinfo {volume} {79}},\ \bibinfo
  {pages} {205432} (\bibinfo {year} {2009})},\ \Eprint
  {http://arxiv.org/abs/0902.4402} {arXiv:0902.4402} \BibitemShut {NoStop}%
\bibitem [{\citenamefont {Kainaris}\ and\ \citenamefont
  {Carr}(2015)}]{Kainaris2015a}%
  \BibitemOpen
  \bibfield  {author} {\bibinfo {author} {\bibfnamefont {N.}~\bibnamefont
  {Kainaris}}\ and\ \bibinfo {author} {\bibfnamefont {S.~T.}\ \bibnamefont
  {Carr}},\ }\href {\doibase 10.1103/PhysRevB.92.035139} {\bibfield  {journal}
  {\bibinfo  {journal} {Phys. Rev. B}\ }\textbf {\bibinfo {volume} {92}},\
  \bibinfo {pages} {035139} (\bibinfo {year} {2015})},\ \Eprint
  {http://arxiv.org/abs/1504.05016} {arXiv:1504.05016} \BibitemShut {NoStop}%
\bibitem [{\citenamefont {Hevroni}\ \emph {et~al.}(2016)\citenamefont
  {Hevroni}, \citenamefont {Shelukhin}, \citenamefont {Karpovski},
  \citenamefont {Goldstein}, \citenamefont {Sela}, \citenamefont {Shtrikman},\
  and\ \citenamefont {Palevski}}]{Hevroni2016}%
  \BibitemOpen
  \bibfield  {author} {\bibinfo {author} {\bibfnamefont {R.}~\bibnamefont
  {Hevroni}}, \bibinfo {author} {\bibfnamefont {V.}~\bibnamefont {Shelukhin}},
  \bibinfo {author} {\bibfnamefont {M.}~\bibnamefont {Karpovski}}, \bibinfo
  {author} {\bibfnamefont {M.}~\bibnamefont {Goldstein}}, \bibinfo {author}
  {\bibfnamefont {E.}~\bibnamefont {Sela}}, \bibinfo {author} {\bibfnamefont
  {H.}~\bibnamefont {Shtrikman}}, \ and\ \bibinfo {author} {\bibfnamefont
  {A.}~\bibnamefont {Palevski}},\ }\href {\doibase 10.1103/PhysRevB.93.035305}
  {\bibfield  {journal} {\bibinfo  {journal} {Phys. Rev. B}\ }\textbf {\bibinfo
  {volume} {93}},\ \bibinfo {pages} {035305} (\bibinfo {year} {2016})},\
  \Eprint {http://arxiv.org/abs/1504.03463} {arXiv:1504.03463} \BibitemShut
  {NoStop}%
\bibitem [{\citenamefont {Heedt}\ \emph {et~al.}(2017)\citenamefont {Heedt},
  \citenamefont {{Traverso Ziani}}, \citenamefont {Cr{\'{e}}pin}, \citenamefont
  {Prost}, \citenamefont {Trellenkamp}, \citenamefont {Schubert}, \citenamefont
  {Gr{\"{u}}tzmacher}, \citenamefont {Trauzettel},\ and\ \citenamefont
  {Sch{\"{a}}pers}}]{Heedt2017}%
  \BibitemOpen
  \bibfield  {author} {\bibinfo {author} {\bibfnamefont {S.}~\bibnamefont
  {Heedt}}, \bibinfo {author} {\bibfnamefont {N.}~\bibnamefont {{Traverso
  Ziani}}}, \bibinfo {author} {\bibfnamefont {F.}~\bibnamefont {Cr{\'{e}}pin}},
  \bibinfo {author} {\bibfnamefont {W.}~\bibnamefont {Prost}}, \bibinfo
  {author} {\bibfnamefont {S.}~\bibnamefont {Trellenkamp}}, \bibinfo {author}
  {\bibfnamefont {J.}~\bibnamefont {Schubert}}, \bibinfo {author}
  {\bibfnamefont {D.}~\bibnamefont {Gr{\"{u}}tzmacher}}, \bibinfo {author}
  {\bibfnamefont {B.}~\bibnamefont {Trauzettel}}, \ and\ \bibinfo {author}
  {\bibfnamefont {T.}~\bibnamefont {Sch{\"{a}}pers}},\ }\href {\doibase
  10.1038/nphys4070} {\bibfield  {journal} {\bibinfo  {journal} {Nat. Phys.}\
  }\textbf {\bibinfo {volume} {13}},\ \bibinfo {pages} {563} (\bibinfo {year}
  {2017})},\ \Eprint {http://arxiv.org/abs/1701.08439} {arXiv:1701.08439}
  \BibitemShut {NoStop}%
\bibitem [{\citenamefont {Tarucha}\ \emph {et~al.}(1995)\citenamefont
  {Tarucha}, \citenamefont {Honda},\ and\ \citenamefont {Saku}}]{Tarucha1995}%
  \BibitemOpen
  \bibfield  {author} {\bibinfo {author} {\bibfnamefont {S.}~\bibnamefont
  {Tarucha}}, \bibinfo {author} {\bibfnamefont {T.}~\bibnamefont {Honda}}, \
  and\ \bibinfo {author} {\bibfnamefont {T.}~\bibnamefont {Saku}},\ }\href
  {\doibase 10.1016/0038-1098(95)00102-6} {\bibfield  {journal} {\bibinfo
  {journal} {Solid State Commun.}\ }\textbf {\bibinfo {volume} {94}},\ \bibinfo
  {pages} {413} (\bibinfo {year} {1995})}\BibitemShut {NoStop}%
\bibitem [{\citenamefont {Asayama}\ \emph {et~al.}(2002)\citenamefont
  {Asayama}, \citenamefont {Tokura}, \citenamefont {Miyashita}, \citenamefont
  {Stopa},\ and\ \citenamefont {Tarucha}}]{Asayama2002}%
  \BibitemOpen
  \bibfield  {author} {\bibinfo {author} {\bibfnamefont {T.}~\bibnamefont
  {Asayama}}, \bibinfo {author} {\bibfnamefont {Y.}~\bibnamefont {Tokura}},
  \bibinfo {author} {\bibfnamefont {S.}~\bibnamefont {Miyashita}}, \bibinfo
  {author} {\bibfnamefont {M.}~\bibnamefont {Stopa}}, \ and\ \bibinfo {author}
  {\bibfnamefont {S.}~\bibnamefont {Tarucha}},\ }\href {\doibase
  10.1016/S1386-9477(01)00302-2} {\bibfield  {journal} {\bibinfo  {journal}
  {Phys. E}\ }\textbf {\bibinfo {volume} {12}},\ \bibinfo {pages} {186}
  (\bibinfo {year} {2002})}\BibitemShut {NoStop}%
\bibitem [{\citenamefont {Levy}\ \emph {et~al.}(2006)\citenamefont {Levy},
  \citenamefont {Tsukernik}, \citenamefont {Karpovski}, \citenamefont
  {Palevski}, \citenamefont {Dwir}, \citenamefont {Pelucchi}, \citenamefont
  {Rudra}, \citenamefont {Kapon},\ and\ \citenamefont {Oreg}}]{Levy2006}%
  \BibitemOpen
  \bibfield  {author} {\bibinfo {author} {\bibfnamefont {E.}~\bibnamefont
  {Levy}}, \bibinfo {author} {\bibfnamefont {A.}~\bibnamefont {Tsukernik}},
  \bibinfo {author} {\bibfnamefont {M.}~\bibnamefont {Karpovski}}, \bibinfo
  {author} {\bibfnamefont {A.}~\bibnamefont {Palevski}}, \bibinfo {author}
  {\bibfnamefont {B.}~\bibnamefont {Dwir}}, \bibinfo {author} {\bibfnamefont
  {E.}~\bibnamefont {Pelucchi}}, \bibinfo {author} {\bibfnamefont
  {A.}~\bibnamefont {Rudra}}, \bibinfo {author} {\bibfnamefont
  {E.}~\bibnamefont {Kapon}}, \ and\ \bibinfo {author} {\bibfnamefont
  {Y.}~\bibnamefont {Oreg}},\ }\href {\doibase 10.1103/PhysRevLett.97.196802}
  {\bibfield  {journal} {\bibinfo  {journal} {Phys. Rev. Lett.}\ }\textbf
  {\bibinfo {volume} {97}},\ \bibinfo {pages} {196802} (\bibinfo {year}
  {2006})},\ \Eprint {http://arxiv.org/abs/0509027} {arXiv:0509027 [cond-mat]}
  \BibitemShut {NoStop}%
\bibitem [{\citenamefont {Levy}\ \emph {et~al.}(2012)\citenamefont {Levy},
  \citenamefont {Sternfeld}, \citenamefont {Eshkol}, \citenamefont {Karpovski},
  \citenamefont {Dwir}, \citenamefont {Rudra}, \citenamefont {Kapon},
  \citenamefont {Oreg},\ and\ \citenamefont {Palevski}}]{Levy2012}%
  \BibitemOpen
  \bibfield  {author} {\bibinfo {author} {\bibfnamefont {E.}~\bibnamefont
  {Levy}}, \bibinfo {author} {\bibfnamefont {I.}~\bibnamefont {Sternfeld}},
  \bibinfo {author} {\bibfnamefont {M.}~\bibnamefont {Eshkol}}, \bibinfo
  {author} {\bibfnamefont {M.}~\bibnamefont {Karpovski}}, \bibinfo {author}
  {\bibfnamefont {B.}~\bibnamefont {Dwir}}, \bibinfo {author} {\bibfnamefont
  {A.}~\bibnamefont {Rudra}}, \bibinfo {author} {\bibfnamefont
  {E.}~\bibnamefont {Kapon}}, \bibinfo {author} {\bibfnamefont
  {Y.}~\bibnamefont {Oreg}}, \ and\ \bibinfo {author} {\bibfnamefont
  {A.}~\bibnamefont {Palevski}},\ }\href {\doibase 10.1103/PhysRevB.85.045315}
  {\bibfield  {journal} {\bibinfo  {journal} {Phys. Rev. B}\ }\textbf {\bibinfo
  {volume} {85}},\ \bibinfo {pages} {045315} (\bibinfo {year}
  {2012})}\BibitemShut {NoStop}%
\bibitem [{\citenamefont {Bockrath}\ \emph {et~al.}(1999)\citenamefont
  {Bockrath}, \citenamefont {Cobden}, \citenamefont {Lu}, \citenamefont
  {Rinzler}, \citenamefont {Smalley}, \citenamefont {Balents},\ and\
  \citenamefont {McEuen}}]{Bockrath1999}%
  \BibitemOpen
  \bibfield  {author} {\bibinfo {author} {\bibfnamefont {M.}~\bibnamefont
  {Bockrath}}, \bibinfo {author} {\bibfnamefont {D.~H.}\ \bibnamefont
  {Cobden}}, \bibinfo {author} {\bibfnamefont {J.}~\bibnamefont {Lu}}, \bibinfo
  {author} {\bibfnamefont {A.~G.}\ \bibnamefont {Rinzler}}, \bibinfo {author}
  {\bibfnamefont {R.~E.}\ \bibnamefont {Smalley}}, \bibinfo {author}
  {\bibfnamefont {L.}~\bibnamefont {Balents}}, \ and\ \bibinfo {author}
  {\bibfnamefont {P.~L.}\ \bibnamefont {McEuen}},\ }\href {\doibase
  10.1038/17569} {\bibfield  {journal} {\bibinfo  {journal} {Nature}\ }\textbf
  {\bibinfo {volume} {397}},\ \bibinfo {pages} {598} (\bibinfo {year}
  {1999})},\ \Eprint {http://arxiv.org/abs/9812233} {arXiv:9812233 [cond-mat]}
  \BibitemShut {NoStop}%
\bibitem [{\citenamefont {Postma}\ \emph {et~al.}(2000)\citenamefont {Postma},
  \citenamefont {de~Jonge}, \citenamefont {Yao},\ and\ \citenamefont
  {Dekker}}]{Postma2000}%
  \BibitemOpen
  \bibfield  {author} {\bibinfo {author} {\bibfnamefont {H.~W.~C.}\
  \bibnamefont {Postma}}, \bibinfo {author} {\bibfnamefont {M.}~\bibnamefont
  {de~Jonge}}, \bibinfo {author} {\bibfnamefont {Z.}~\bibnamefont {Yao}}, \
  and\ \bibinfo {author} {\bibfnamefont {C.}~\bibnamefont {Dekker}},\ }\href
  {\doibase 10.1103/PhysRevB.62.R10653} {\bibfield  {journal} {\bibinfo
  {journal} {Phys. Rev. B}\ }\textbf {\bibinfo {volume} {62}},\ \bibinfo
  {pages} {R10653} (\bibinfo {year} {2000})},\ \Eprint
  {http://arxiv.org/abs/0009055} {arXiv:0009055 [cond-mat]} \BibitemShut
  {NoStop}%
\bibitem [{\citenamefont {Graugnard}\ \emph {et~al.}(2001)\citenamefont
  {Graugnard}, \citenamefont {de~Pablo}, \citenamefont {Walsh}, \citenamefont
  {Ghosh}, \citenamefont {Datta},\ and\ \citenamefont
  {Reifenberger}}]{Graugnard2001}%
  \BibitemOpen
  \bibfield  {author} {\bibinfo {author} {\bibfnamefont {E.}~\bibnamefont
  {Graugnard}}, \bibinfo {author} {\bibfnamefont {P.~J.}\ \bibnamefont
  {de~Pablo}}, \bibinfo {author} {\bibfnamefont {B.}~\bibnamefont {Walsh}},
  \bibinfo {author} {\bibfnamefont {A.~W.}\ \bibnamefont {Ghosh}}, \bibinfo
  {author} {\bibfnamefont {S.}~\bibnamefont {Datta}}, \ and\ \bibinfo {author}
  {\bibfnamefont {R.}~\bibnamefont {Reifenberger}},\ }\href {\doibase
  10.1103/PhysRevB.64.125407} {\bibfield  {journal} {\bibinfo  {journal} {Phys.
  Rev. B}\ }\textbf {\bibinfo {volume} {64}},\ \bibinfo {pages} {125407}
  (\bibinfo {year} {2001})}\BibitemShut {NoStop}%
\bibitem [{\citenamefont {Bachtold}\ \emph {et~al.}(2001)\citenamefont
  {Bachtold}, \citenamefont {de~Jonge}, \citenamefont {Grove-Rasmussen},
  \citenamefont {McEuen}, \citenamefont {Buitelaar},\ and\ \citenamefont
  {Sch{\"{o}}nenberger}}]{Bachtold2001}%
  \BibitemOpen
  \bibfield  {author} {\bibinfo {author} {\bibfnamefont {A.}~\bibnamefont
  {Bachtold}}, \bibinfo {author} {\bibfnamefont {M.}~\bibnamefont {de~Jonge}},
  \bibinfo {author} {\bibfnamefont {K.}~\bibnamefont {Grove-Rasmussen}},
  \bibinfo {author} {\bibfnamefont {P.~L.}\ \bibnamefont {McEuen}}, \bibinfo
  {author} {\bibfnamefont {M.}~\bibnamefont {Buitelaar}}, \ and\ \bibinfo
  {author} {\bibfnamefont {C.}~\bibnamefont {Sch{\"{o}}nenberger}},\ }\href
  {\doibase 10.1103/PhysRevLett.87.166801} {\bibfield  {journal} {\bibinfo
  {journal} {Phys. Rev. Lett.}\ }\textbf {\bibinfo {volume} {87}},\ \bibinfo
  {pages} {166801} (\bibinfo {year} {2001})},\ \Eprint
  {http://arxiv.org/abs/0012262} {arXiv:0012262 [cond-mat]} \BibitemShut
  {NoStop}%
\bibitem [{\citenamefont {Liu}\ \emph {et~al.}(2001)\citenamefont {Liu},
  \citenamefont {Avouris}, \citenamefont {Martel},\ and\ \citenamefont
  {Hsu}}]{Liu2001}%
  \BibitemOpen
  \bibfield  {author} {\bibinfo {author} {\bibfnamefont {K.}~\bibnamefont
  {Liu}}, \bibinfo {author} {\bibfnamefont {P.}~\bibnamefont {Avouris}},
  \bibinfo {author} {\bibfnamefont {R.}~\bibnamefont {Martel}}, \ and\ \bibinfo
  {author} {\bibfnamefont {W.~K.}\ \bibnamefont {Hsu}},\ }\href {\doibase
  10.1103/PhysRevB.63.161404} {\bibfield  {journal} {\bibinfo  {journal} {Phys.
  Rev. B}\ }\textbf {\bibinfo {volume} {63}},\ \bibinfo {pages} {161404}
  (\bibinfo {year} {2001})}\BibitemShut {NoStop}%
\bibitem [{\citenamefont {Lee}\ \emph {et~al.}(2004)\citenamefont {Lee},
  \citenamefont {Eggert}, \citenamefont {Kim}, \citenamefont {Kahng},
  \citenamefont {Shinohara},\ and\ \citenamefont {Kuk}}]{Lee2004}%
  \BibitemOpen
  \bibfield  {author} {\bibinfo {author} {\bibfnamefont {J.}~\bibnamefont
  {Lee}}, \bibinfo {author} {\bibfnamefont {S.}~\bibnamefont {Eggert}},
  \bibinfo {author} {\bibfnamefont {H.}~\bibnamefont {Kim}}, \bibinfo {author}
  {\bibfnamefont {S.-J.}\ \bibnamefont {Kahng}}, \bibinfo {author}
  {\bibfnamefont {H.}~\bibnamefont {Shinohara}}, \ and\ \bibinfo {author}
  {\bibfnamefont {Y.}~\bibnamefont {Kuk}},\ }\href {\doibase
  10.1103/PhysRevLett.93.166403} {\bibfield  {journal} {\bibinfo  {journal}
  {Phys. Rev. Lett.}\ }\textbf {\bibinfo {volume} {93}},\ \bibinfo {pages}
  {166403} (\bibinfo {year} {2004})},\ \Eprint {http://arxiv.org/abs/0412001}
  {arXiv:0412001 [cond-mat]} \BibitemShut {NoStop}%
\bibitem [{\citenamefont {Anthore}\ \emph {et~al.}(2018)\citenamefont
  {Anthore}, \citenamefont {Iftikhar}, \citenamefont {Boulat}, \citenamefont
  {Parmentier}, \citenamefont {Cavanna}, \citenamefont {Ouerghi}, \citenamefont
  {Gennser},\ and\ \citenamefont {Pierre}}]{Anthore2018}%
  \BibitemOpen
  \bibfield  {author} {\bibinfo {author} {\bibfnamefont {A.}~\bibnamefont
  {Anthore}}, \bibinfo {author} {\bibfnamefont {Z.}~\bibnamefont {Iftikhar}},
  \bibinfo {author} {\bibfnamefont {E.}~\bibnamefont {Boulat}}, \bibinfo
  {author} {\bibfnamefont {F.~D.}\ \bibnamefont {Parmentier}}, \bibinfo
  {author} {\bibfnamefont {A.}~\bibnamefont {Cavanna}}, \bibinfo {author}
  {\bibfnamefont {A.}~\bibnamefont {Ouerghi}}, \bibinfo {author} {\bibfnamefont
  {U.}~\bibnamefont {Gennser}}, \ and\ \bibinfo {author} {\bibfnamefont
  {F.}~\bibnamefont {Pierre}},\ }\href {\doibase 10.1103/PhysRevX.8.031075}
  {\bibfield  {journal} {\bibinfo  {journal} {Phys. Rev. X}\ }\textbf {\bibinfo
  {volume} {8}},\ \bibinfo {pages} {031075} (\bibinfo {year} {2018})},\ \Eprint
  {http://arxiv.org/abs/1809.02017} {arXiv:1809.02017} \BibitemShut {NoStop}%
\bibitem [{\citenamefont {Luo}\ \emph {et~al.}(1990)\citenamefont {Luo},
  \citenamefont {Munekata}, \citenamefont {Fang},\ and\ \citenamefont
  {Stiles}}]{Luo1990}%
  \BibitemOpen
  \bibfield  {author} {\bibinfo {author} {\bibfnamefont {J.}~\bibnamefont
  {Luo}}, \bibinfo {author} {\bibfnamefont {H.}~\bibnamefont {Munekata}},
  \bibinfo {author} {\bibfnamefont {F.~F.}\ \bibnamefont {Fang}}, \ and\
  \bibinfo {author} {\bibfnamefont {P.~J.}\ \bibnamefont {Stiles}},\ }\href
  {\doibase 10.1103/PhysRevB.41.7685} {\bibfield  {journal} {\bibinfo
  {journal} {Phys. Rev. B}\ }\textbf {\bibinfo {volume} {41}},\ \bibinfo
  {pages} {7685} (\bibinfo {year} {1990})}\BibitemShut {NoStop}%
\bibitem [{\citenamefont {Matsuo}\ \emph {et~al.}(2017)\citenamefont {Matsuo},
  \citenamefont {Kamata}, \citenamefont {Baba}, \citenamefont {Deacon},
  \citenamefont {Shabani}, \citenamefont {Palmstr{\o}m},\ and\ \citenamefont
  {Tarucha}}]{Matsuo2017}%
  \BibitemOpen
  \bibfield  {author} {\bibinfo {author} {\bibfnamefont {S.}~\bibnamefont
  {Matsuo}}, \bibinfo {author} {\bibfnamefont {H.}~\bibnamefont {Kamata}},
  \bibinfo {author} {\bibfnamefont {S.}~\bibnamefont {Baba}}, \bibinfo {author}
  {\bibfnamefont {R.~S.}\ \bibnamefont {Deacon}}, \bibinfo {author}
  {\bibfnamefont {J.}~\bibnamefont {Shabani}}, \bibinfo {author} {\bibfnamefont
  {C.~J.}\ \bibnamefont {Palmstr{\o}m}}, \ and\ \bibinfo {author}
  {\bibfnamefont {S.}~\bibnamefont {Tarucha}},\ }\href {\doibase
  10.1103/PhysRevB.96.201404} {\bibfield  {journal} {\bibinfo  {journal} {Phys.
  Rev. B}\ }\textbf {\bibinfo {volume} {96}},\ \bibinfo {pages} {201404}
  (\bibinfo {year} {2017})},\ \Eprint {http://arxiv.org/abs/1707.09515}
  {arXiv:1707.09515} \BibitemShut {NoStop}%
\bibitem [{\citenamefont {Balents}(1999)}]{Balents1999}%
  \BibitemOpen
  \bibfield  {author} {\bibinfo {author} {\bibfnamefont {L.}~\bibnamefont
  {Balents}},\ }\href {\doibase 00178012} {\bibfield  {journal} {\bibinfo
  {journal} {arXiv Prepr. cond-mat/9906032}\ ,\ \bibinfo {pages} {7}} (\bibinfo
  {year} {1999})},\ \Eprint {http://arxiv.org/abs/9906032} {arXiv:9906032
  [cond-mat]} \BibitemShut {NoStop}%
\bibitem [{\citenamefont {Venkataraman}\ \emph {et~al.}(2006)\citenamefont
  {Venkataraman}, \citenamefont {Hong},\ and\ \citenamefont
  {Kim}}]{Venkataraman2006}%
  \BibitemOpen
  \bibfield  {author} {\bibinfo {author} {\bibfnamefont {L.}~\bibnamefont
  {Venkataraman}}, \bibinfo {author} {\bibfnamefont {Y.~S.}\ \bibnamefont
  {Hong}}, \ and\ \bibinfo {author} {\bibfnamefont {P.}~\bibnamefont {Kim}},\
  }\href {\doibase 10.1103/PhysRevLett.96.076601} {\bibfield  {journal}
  {\bibinfo  {journal} {Phys. Rev. Lett.}\ }\textbf {\bibinfo {volume} {96}},\
  \bibinfo {pages} {076601} (\bibinfo {year} {2006})},\ \Eprint
  {http://arxiv.org/abs/0601454} {arXiv:0601454 [cond-mat]} \BibitemShut
  {NoStop}%
\bibitem [{\citenamefont {Hsu}\ \emph {et~al.}()\citenamefont {Hsu},
  \citenamefont {Stano}, \citenamefont {Sato}, \citenamefont {Matsuo},
  \citenamefont {Tarucha},\ and\ \citenamefont {Loss}}]{Hsu2018b}%
  \BibitemOpen
  \bibfield  {author} {\bibinfo {author} {\bibfnamefont {C.-H.}\ \bibnamefont
  {Hsu}}, \bibinfo {author} {\bibfnamefont {P.}~\bibnamefont {Stano}}, \bibinfo
  {author} {\bibfnamefont {Y.}~\bibnamefont {Sato}}, \bibinfo {author}
  {\bibfnamefont {S.}~\bibnamefont {Matsuo}}, \bibinfo {author} {\bibfnamefont
  {S.}~\bibnamefont {Tarucha}}, \ and\ \bibinfo {author} {\bibfnamefont
  {D.}~\bibnamefont {Loss}},\ }\href@noop {} {\ }\bibinfo {note}
  {(unpublished)}\BibitemShut {NoStop}%
\bibitem [{\citenamefont {Chang}\ \emph {et~al.}(1996)\citenamefont {Chang},
  \citenamefont {Pfeiffer},\ and\ \citenamefont {West}}]{Chang1996}%
  \BibitemOpen
  \bibfield  {author} {\bibinfo {author} {\bibfnamefont {A.~M.}\ \bibnamefont
  {Chang}}, \bibinfo {author} {\bibfnamefont {L.~N.}\ \bibnamefont {Pfeiffer}},
  \ and\ \bibinfo {author} {\bibfnamefont {K.~W.}\ \bibnamefont {West}},\
  }\href {\doibase 10.1103/PhysRevLett.77.2538} {\bibfield  {journal} {\bibinfo
   {journal} {Phys. Rev. Lett.}\ }\textbf {\bibinfo {volume} {77}},\ \bibinfo
  {pages} {2538} (\bibinfo {year} {1996})}\BibitemShut {NoStop}%
\bibitem [{\citenamefont {Slot}\ \emph {et~al.}(2004)\citenamefont {Slot},
  \citenamefont {Holst}, \citenamefont {van~der Zant},\ and\ \citenamefont
  {Zaitsev-Zotov}}]{Slot2004}%
  \BibitemOpen
  \bibfield  {author} {\bibinfo {author} {\bibfnamefont {E.}~\bibnamefont
  {Slot}}, \bibinfo {author} {\bibfnamefont {M.~A.}\ \bibnamefont {Holst}},
  \bibinfo {author} {\bibfnamefont {H.~S.~J.}\ \bibnamefont {van~der Zant}}, \
  and\ \bibinfo {author} {\bibfnamefont {S.~V.}\ \bibnamefont
  {Zaitsev-Zotov}},\ }\href {\doibase 10.1103/PhysRevLett.93.176602} {\bibfield
   {journal} {\bibinfo  {journal} {Phys. Rev. Lett.}\ }\textbf {\bibinfo
  {volume} {93}},\ \bibinfo {pages} {176602} (\bibinfo {year}
  {2004})}\BibitemShut {NoStop}%
\bibitem [{\citenamefont {Aleshin}\ \emph {et~al.}(2004)\citenamefont
  {Aleshin}, \citenamefont {Lee}, \citenamefont {Park},\ and\ \citenamefont
  {Akagi}}]{Aleshin2004}%
  \BibitemOpen
  \bibfield  {author} {\bibinfo {author} {\bibfnamefont {A.~N.}\ \bibnamefont
  {Aleshin}}, \bibinfo {author} {\bibfnamefont {H.~J.}\ \bibnamefont {Lee}},
  \bibinfo {author} {\bibfnamefont {Y.~W.}\ \bibnamefont {Park}}, \ and\
  \bibinfo {author} {\bibfnamefont {K.}~\bibnamefont {Akagi}},\ }\href
  {\doibase 10.1103/PhysRevLett.93.196601} {\bibfield  {journal} {\bibinfo
  {journal} {Phys. Rev. Lett.}\ }\textbf {\bibinfo {volume} {93}},\ \bibinfo
  {pages} {196601} (\bibinfo {year} {2004})}\BibitemShut {NoStop}%
\bibitem [{\citenamefont {Lucot}\ \emph {et~al.}(2011)\citenamefont {Lucot},
  \citenamefont {Jabeen}, \citenamefont {Harmand}, \citenamefont {Patriarche},
  \citenamefont {Giraud}, \citenamefont {Faini},\ and\ \citenamefont
  {Mailly}}]{Lucot2011}%
  \BibitemOpen
  \bibfield  {author} {\bibinfo {author} {\bibfnamefont {D.}~\bibnamefont
  {Lucot}}, \bibinfo {author} {\bibfnamefont {F.}~\bibnamefont {Jabeen}},
  \bibinfo {author} {\bibfnamefont {J.-C.}\ \bibnamefont {Harmand}}, \bibinfo
  {author} {\bibfnamefont {G.}~\bibnamefont {Patriarche}}, \bibinfo {author}
  {\bibfnamefont {R.}~\bibnamefont {Giraud}}, \bibinfo {author} {\bibfnamefont
  {G.}~\bibnamefont {Faini}}, \ and\ \bibinfo {author} {\bibfnamefont
  {D.}~\bibnamefont {Mailly}},\ }\href {\doibase 10.1063/1.3574026} {\bibfield
  {journal} {\bibinfo  {journal} {Appl. Phys. Lett.}\ }\textbf {\bibinfo
  {volume} {98}},\ \bibinfo {pages} {142114} (\bibinfo {year} {2011})},\
  \Eprint {http://arxiv.org/abs/arXiv:1101.0421v1} {arXiv:arXiv:1101.0421v1}
  \BibitemShut {NoStop}%
\bibitem [{\citenamefont {Li}\ \emph {et~al.}(2015)\citenamefont {Li},
  \citenamefont {Wang}, \citenamefont {Fu}, \citenamefont {Du}, \citenamefont
  {Schreiber}, \citenamefont {Mu}, \citenamefont {Liu}, \citenamefont
  {Sullivan}, \citenamefont {Cs{\'{a}}thy}, \citenamefont {Lin},\ and\
  \citenamefont {Du}}]{Li2015}%
  \BibitemOpen
  \bibfield  {author} {\bibinfo {author} {\bibfnamefont {T.}~\bibnamefont
  {Li}}, \bibinfo {author} {\bibfnamefont {P.}~\bibnamefont {Wang}}, \bibinfo
  {author} {\bibfnamefont {H.}~\bibnamefont {Fu}}, \bibinfo {author}
  {\bibfnamefont {L.}~\bibnamefont {Du}}, \bibinfo {author} {\bibfnamefont
  {K.~A.}\ \bibnamefont {Schreiber}}, \bibinfo {author} {\bibfnamefont
  {X.}~\bibnamefont {Mu}}, \bibinfo {author} {\bibfnamefont {X.}~\bibnamefont
  {Liu}}, \bibinfo {author} {\bibfnamefont {G.}~\bibnamefont {Sullivan}},
  \bibinfo {author} {\bibfnamefont {G.~A.}\ \bibnamefont {Cs{\'{a}}thy}},
  \bibinfo {author} {\bibfnamefont {X.}~\bibnamefont {Lin}}, \ and\ \bibinfo
  {author} {\bibfnamefont {R.-R.}\ \bibnamefont {Du}},\ }\href {\doibase
  10.1103/PhysRevLett.115.136804} {\bibfield  {journal} {\bibinfo  {journal}
  {Phys. Rev. Lett.}\ }\textbf {\bibinfo {volume} {115}},\ \bibinfo {pages}
  {136804} (\bibinfo {year} {2015})}\BibitemShut {NoStop}%
\bibitem [{\citenamefont {Moroz}\ and\ \citenamefont
  {Barnes}(1999)}]{Moroz1999}%
  \BibitemOpen
  \bibfield  {author} {\bibinfo {author} {\bibfnamefont {A.~V.}\ \bibnamefont
  {Moroz}}\ and\ \bibinfo {author} {\bibfnamefont {C.~H.~W.}\ \bibnamefont
  {Barnes}},\ }\href {\doibase 10.1103/PhysRevB.60.14272} {\bibfield  {journal}
  {\bibinfo  {journal} {Phys. Rev. B}\ }\textbf {\bibinfo {volume} {60}},\
  \bibinfo {pages} {14272} (\bibinfo {year} {1999})},\ \Eprint
  {http://arxiv.org/abs/9910466} {arXiv:9910466 [cond-mat]} \BibitemShut
  {NoStop}%
\bibitem [{Note1()}]{Note1}%
  \BibitemOpen
  \bibinfo {note} {As the wires are much longer than the bulk mean free path of
  \SI {690}{nm}, the disorder (perhaps, in the form of weak potential
  modulation due to impurities) should play role for the wire
  resistance.}\BibitemShut {Stop}%
\bibitem [{Note2()}]{Note2}%
  \BibitemOpen
  \bibinfo {note} {One may consider a third scenario, in which one of barriers
  is in the bulk and the other is at a boundary. In contrast to our
  observation, however, it would give different scaling behavior in the
  high-bias and high-temperature regimes~\cite {Venkataraman2006}. We therefore
  believe that this scenario is not relevant to our data.}\BibitemShut {Stop}%
\bibitem [{\citenamefont {Kane}\ and\ \citenamefont
  {Fisher}(1992{\natexlab{a}})}]{Kane1992}%
  \BibitemOpen
  \bibfield  {author} {\bibinfo {author} {\bibfnamefont {C.~L.}\ \bibnamefont
  {Kane}}\ and\ \bibinfo {author} {\bibfnamefont {M.~P.~A.}\ \bibnamefont
  {Fisher}},\ }\href {\doibase 10.1103/PhysRevB.46.15233} {\bibfield  {journal}
  {\bibinfo  {journal} {Phys. Rev. B}\ }\textbf {\bibinfo {volume} {46}},\
  \bibinfo {pages} {15233} (\bibinfo {year} {1992}{\natexlab{a}})}\BibitemShut
  {NoStop}%
\bibitem [{\citenamefont {Giamarchi}(2003)}]{Giamarchi2003a}%
  \BibitemOpen
  \bibfield  {author} {\bibinfo {author} {\bibfnamefont {T.}~\bibnamefont
  {Giamarchi}},\ }\href {\doibase 10.1093/acprof:oso/9780198525004.001.0001}
  {\emph {\bibinfo {title} {{Quantum Physics in One Dimension}}}}\ (\bibinfo
  {publisher} {Oxford University Press},\ \bibinfo {year} {2003})\BibitemShut
  {NoStop}%
\bibitem [{\citenamefont {Maciejko}\ \emph {et~al.}(2009)\citenamefont
  {Maciejko}, \citenamefont {Liu}, \citenamefont {Oreg}, \citenamefont {Qi},
  \citenamefont {Wu},\ and\ \citenamefont {Zhang}}]{Maciejko2009}%
  \BibitemOpen
  \bibfield  {author} {\bibinfo {author} {\bibfnamefont {J.}~\bibnamefont
  {Maciejko}}, \bibinfo {author} {\bibfnamefont {C.}~\bibnamefont {Liu}},
  \bibinfo {author} {\bibfnamefont {Y.}~\bibnamefont {Oreg}}, \bibinfo {author}
  {\bibfnamefont {X.-L.}\ \bibnamefont {Qi}}, \bibinfo {author} {\bibfnamefont
  {C.}~\bibnamefont {Wu}}, \ and\ \bibinfo {author} {\bibfnamefont {S.-C.}\
  \bibnamefont {Zhang}},\ }\href {\doibase 10.1103/PhysRevLett.102.256803}
  {\bibfield  {journal} {\bibinfo  {journal} {Phys. Rev. Lett.}\ }\textbf
  {\bibinfo {volume} {102}},\ \bibinfo {pages} {256803} (\bibinfo {year}
  {2009})},\ \Eprint {http://arxiv.org/abs/0901.1685} {arXiv:0901.1685}
  \BibitemShut {NoStop}%
\bibitem [{\citenamefont {Levinshtein}\ \emph {et~al.}(1996)\citenamefont
  {Levinshtein}, \citenamefont {Rumyantsev},\ and\ \citenamefont
  {Shur}}]{Levinshtein1996}%
  \BibitemOpen
  \bibfield  {author} {\bibinfo {author} {\bibfnamefont {M.}~\bibnamefont
  {Levinshtein}}, \bibinfo {author} {\bibfnamefont {S.}~\bibnamefont
  {Rumyantsev}}, \ and\ \bibinfo {author} {\bibfnamefont {M.}~\bibnamefont
  {Shur}},\ }\href {\doibase 10.1142/2046-vol1} {\emph {\bibinfo {title}
  {Handb. Ser. Semicond. Parameters}}},\ Vol.~\bibinfo {volume} {1}\ (\bibinfo
  {publisher} {WORLD SCIENTIFIC},\ \bibinfo {year} {1996})\BibitemShut
  {NoStop}%
\bibitem [{Note3()}]{Note3}%
  \BibitemOpen
  \bibinfo {note} {From the stacking structure of the wafer, we estimate the
  Fermi energy $E_\protect \textrm {F}=1.13\times 10^2\times [V_g-(-0.86)]^2$
  [\si {m\electronvolt }] and the Fermi velocity $v_\protect \textrm
  {F}=1.31\times 10^{13}\times [V_g-(-0.86)]w$ [\si {m/s}][See appendix]. Here,
  $V_g$ is in units of \si {V}, $w$ is in units of \si {m}.}\BibitemShut
  {Stop}%
\bibitem [{Note4()}]{Note4}%
  \BibitemOpen
  \bibinfo {note} {In addition, the level spacing is large enough so we can
  ignore higher subbands at temperatures of our measurements. In Ref.~\cite
  {Heedt2016}, the subband level spacing in a \SI {100}{nm}-diameter InAs
  nanowire with isotropic cross-section of about \SI {8}{m\electronvolt } was
  found, similar to our estimate here.}\BibitemShut {Stop}%
\bibitem [{Note5()}]{Note5}%
  \BibitemOpen
  \bibinfo {note} {If they originate in random disorder, there is no reason for
  such uniformity. On the other hand, one could argue that disorder average
  over many parallel wires might result in a scaling curve with some effective
  number of tunnel barriers, being here close to 2. However, performing such a
  calculation would require to adopt some ad-hoc assumptions about the
  statistical distributions of the strength and position of the tunnel
  barriers. We therefore do not follow this idea.}\BibitemShut {Stop}%
\bibitem [{\citenamefont {Rother}\ \emph {et~al.}(2000)\citenamefont {Rother},
  \citenamefont {Wegscheider}, \citenamefont {Deutschmann}, \citenamefont
  {Bichler},\ and\ \citenamefont {Abstreiter}}]{Rother2000}%
  \BibitemOpen
  \bibfield  {author} {\bibinfo {author} {\bibfnamefont {M.}~\bibnamefont
  {Rother}}, \bibinfo {author} {\bibfnamefont {W.}~\bibnamefont {Wegscheider}},
  \bibinfo {author} {\bibfnamefont {R.}~\bibnamefont {Deutschmann}}, \bibinfo
  {author} {\bibfnamefont {M.}~\bibnamefont {Bichler}}, \ and\ \bibinfo
  {author} {\bibfnamefont {G.}~\bibnamefont {Abstreiter}},\ }\href {\doibase
  10.1016/S1386-9477(99)00106-X} {\bibfield  {journal} {\bibinfo  {journal}
  {Phys. E}\ }\textbf {\bibinfo {volume} {6}},\ \bibinfo {pages} {551}
  (\bibinfo {year} {2000})}\BibitemShut {NoStop}%
\bibitem [{\citenamefont {Auslaender}\ \emph {et~al.}(2000)\citenamefont
  {Auslaender}, \citenamefont {Yacoby}, \citenamefont {de~Picciotto},
  \citenamefont {Baldwin}, \citenamefont {Pfeiffer},\ and\ \citenamefont
  {West}}]{Auslaender2000}%
  \BibitemOpen
  \bibfield  {author} {\bibinfo {author} {\bibfnamefont {O.~M.}\ \bibnamefont
  {Auslaender}}, \bibinfo {author} {\bibfnamefont {A.}~\bibnamefont {Yacoby}},
  \bibinfo {author} {\bibfnamefont {R.}~\bibnamefont {de~Picciotto}}, \bibinfo
  {author} {\bibfnamefont {K.~W.}\ \bibnamefont {Baldwin}}, \bibinfo {author}
  {\bibfnamefont {L.~N.}\ \bibnamefont {Pfeiffer}}, \ and\ \bibinfo {author}
  {\bibfnamefont {K.~W.}\ \bibnamefont {West}},\ }\href {\doibase
  10.1103/PhysRevLett.84.1764} {\bibfield  {journal} {\bibinfo  {journal}
  {Phys. Rev. Lett.}\ }\textbf {\bibinfo {volume} {84}},\ \bibinfo {pages}
  {1764} (\bibinfo {year} {2000})}\BibitemShut {NoStop}%
\bibitem [{Note6()}]{Note6}%
  \BibitemOpen
  \bibinfo {note} {In the TLL model that we work with here, the constants $g$
  are the only parameters defining the strength of the electron-electron
  interactions. The value of the Fermi velocity, or the relation of the kinetic
  to interaction energies, would also need to be considered to judge the
  ``strength'' of the interactions in a broader context. Here we do not
  consider such implications and mean the statements on the e-e interaction
  strength as solely the statements on the value of constants $g$.}\BibitemShut
  {Stop}%
\bibitem [{\citenamefont {Li}\ and\ \citenamefont {Zhang}(2015)}]{Li2015a}%
  \BibitemOpen
  \bibfield  {author} {\bibinfo {author} {\bibfnamefont {S.}~\bibnamefont
  {Li}}\ and\ \bibinfo {author} {\bibfnamefont {Q.}~\bibnamefont {Zhang}},\
  }\href {\doibase 10.1039/C5RA00848D} {\bibfield  {journal} {\bibinfo
  {journal} {RSC Adv.}\ }\textbf {\bibinfo {volume} {5}},\ \bibinfo {pages}
  {28980} (\bibinfo {year} {2015})}\BibitemShut {NoStop}%
\bibitem [{\citenamefont {Littlejohn}\ \emph {et~al.}(1993)\citenamefont
  {Littlejohn}, \citenamefont {Kim},\ and\ \citenamefont
  {Tian}}]{Littlejohn1993}%
  \BibitemOpen
  \bibfield  {author} {\bibinfo {author} {\bibfnamefont {M.~A.}\ \bibnamefont
  {Littlejohn}}, \bibinfo {author} {\bibfnamefont {K.~W.}\ \bibnamefont {Kim}},
  \ and\ \bibinfo {author} {\bibfnamefont {H.}~\bibnamefont {Tian}},\
  }\href@noop {} {\emph {\bibinfo {title} {Properites Lattice-Matched Strained
  Indium Gall. Arsenide}}}\ (\bibinfo  {publisher} {INSPEC},\ \bibinfo {year}
  {1993})\ pp.\ \bibinfo {pages} {107--116}\BibitemShut {NoStop}%
\bibitem [{\citenamefont {Braunecker}\ \emph {et~al.}(2010)\citenamefont
  {Braunecker}, \citenamefont {Japaridze}, \citenamefont {Klinovaja},\ and\
  \citenamefont {Loss}}]{Braunecker2010}%
  \BibitemOpen
  \bibfield  {author} {\bibinfo {author} {\bibfnamefont {B.}~\bibnamefont
  {Braunecker}}, \bibinfo {author} {\bibfnamefont {G.~I.}\ \bibnamefont
  {Japaridze}}, \bibinfo {author} {\bibfnamefont {J.}~\bibnamefont
  {Klinovaja}}, \ and\ \bibinfo {author} {\bibfnamefont {D.}~\bibnamefont
  {Loss}},\ }\href {\doibase 10.1103/PhysRevB.82.045127} {\bibfield  {journal}
  {\bibinfo  {journal} {Phys. Rev. B}\ }\textbf {\bibinfo {volume} {82}},\
  \bibinfo {pages} {045127} (\bibinfo {year} {2010})},\ \Eprint
  {http://arxiv.org/abs/1004.0467} {arXiv:1004.0467} \BibitemShut {NoStop}%
\bibitem [{\citenamefont {Meng}\ \emph {et~al.}(2014)\citenamefont {Meng},
  \citenamefont {Klinovaja},\ and\ \citenamefont {Loss}}]{Meng2014}%
  \BibitemOpen
  \bibfield  {author} {\bibinfo {author} {\bibfnamefont {T.}~\bibnamefont
  {Meng}}, \bibinfo {author} {\bibfnamefont {J.}~\bibnamefont {Klinovaja}}, \
  and\ \bibinfo {author} {\bibfnamefont {D.}~\bibnamefont {Loss}},\ }\href
  {\doibase 10.1103/PhysRevB.89.205133} {\bibfield  {journal} {\bibinfo
  {journal} {Phys. Rev. B}\ }\textbf {\bibinfo {volume} {89}},\ \bibinfo
  {pages} {205133} (\bibinfo {year} {2014})},\ \Eprint
  {http://arxiv.org/abs/arXiv:1403.2759v1} {arXiv:arXiv:1403.2759v1}
  \BibitemShut {NoStop}%
\bibitem [{\citenamefont {Kane}\ and\ \citenamefont
  {Fisher}(1992{\natexlab{b}})}]{Kane1992a}%
  \BibitemOpen
  \bibfield  {author} {\bibinfo {author} {\bibfnamefont {C.~L.}\ \bibnamefont
  {Kane}}\ and\ \bibinfo {author} {\bibfnamefont {M.~P.~A.}\ \bibnamefont
  {Fisher}},\ }\href {\doibase 10.1103/PhysRevLett.68.1220} {\bibfield
  {journal} {\bibinfo  {journal} {Phys. Rev. Lett.}\ }\textbf {\bibinfo
  {volume} {68}},\ \bibinfo {pages} {1220} (\bibinfo {year}
  {1992}{\natexlab{b}})}\BibitemShut {NoStop}%
\bibitem [{\citenamefont {Maslov}(1995)}]{Maslov1995}%
  \BibitemOpen
  \bibfield  {author} {\bibinfo {author} {\bibfnamefont {D.~L.}\ \bibnamefont
  {Maslov}},\ }\href {\doibase 10.1103/PhysRevB.52.R14368} {\bibfield
  {journal} {\bibinfo  {journal} {Phys. Rev. B}\ }\textbf {\bibinfo {volume}
  {52}},\ \bibinfo {pages} {R14368} (\bibinfo {year} {1995})},\ \Eprint
  {http://arxiv.org/abs/9507119} {arXiv:9507119 [cond-mat]} \BibitemShut
  {NoStop}%
\bibitem [{\citenamefont {Matveev}\ and\ \citenamefont
  {Glazman}(1993)}]{Matveev1993a}%
  \BibitemOpen
  \bibfield  {author} {\bibinfo {author} {\bibfnamefont {K.~A.}\ \bibnamefont
  {Matveev}}\ and\ \bibinfo {author} {\bibfnamefont {L.~I.}\ \bibnamefont
  {Glazman}},\ }\href {\doibase 10.1103/PhysRevLett.70.990} {\bibfield
  {journal} {\bibinfo  {journal} {Phys. Rev. Lett.}\ }\textbf {\bibinfo
  {volume} {70}},\ \bibinfo {pages} {990} (\bibinfo {year} {1993})},\ \Eprint
  {http://arxiv.org/abs/arXiv:1011.1669v3} {arXiv:arXiv:1011.1669v3}
  \BibitemShut {NoStop}%
\bibitem [{\citenamefont {Sandler}\ and\ \citenamefont
  {Maslov}(1997)}]{Sandler1997}%
  \BibitemOpen
  \bibfield  {author} {\bibinfo {author} {\bibfnamefont {N.~P.}\ \bibnamefont
  {Sandler}}\ and\ \bibinfo {author} {\bibfnamefont {D.~L.}\ \bibnamefont
  {Maslov}},\ }\href {\doibase 10.1103/PhysRevB.55.13808} {\bibfield  {journal}
  {\bibinfo  {journal} {Phys. Rev. B}\ }\textbf {\bibinfo {volume} {55}},\
  \bibinfo {pages} {13808} (\bibinfo {year} {1997})},\ \Eprint
  {http://arxiv.org/abs/9701155v1} {arXiv:9701155v1 [arXiv:cond-mat]}
  \BibitemShut {NoStop}%
\bibitem [{\citenamefont {Yao}\ \emph {et~al.}(1999)\citenamefont {Yao},
  \citenamefont {Postma}, \citenamefont {Balents},\ and\ \citenamefont
  {Dekker}}]{Yao1999}%
  \BibitemOpen
  \bibfield  {author} {\bibinfo {author} {\bibfnamefont {Z.}~\bibnamefont
  {Yao}}, \bibinfo {author} {\bibfnamefont {H.~W.~C.}\ \bibnamefont {Postma}},
  \bibinfo {author} {\bibfnamefont {L.}~\bibnamefont {Balents}}, \ and\
  \bibinfo {author} {\bibfnamefont {C.}~\bibnamefont {Dekker}},\ }\href
  {\doibase 10.1038/46241} {\bibfield  {journal} {\bibinfo  {journal} {Nature}\
  }\textbf {\bibinfo {volume} {402}},\ \bibinfo {pages} {273} (\bibinfo {year}
  {1999})}\BibitemShut {NoStop}%
\bibitem [{\citenamefont {Egger}(1999)}]{Egger1999}%
  \BibitemOpen
  \bibfield  {author} {\bibinfo {author} {\bibfnamefont {R.}~\bibnamefont
  {Egger}},\ }\href {\doibase 10.1103/PhysRevLett.83.5547} {\bibfield
  {journal} {\bibinfo  {journal} {Phys. Rev. Lett.}\ }\textbf {\bibinfo
  {volume} {83}},\ \bibinfo {pages} {5547} (\bibinfo {year}
  {1999})}\BibitemShut {NoStop}%
\bibitem [{\citenamefont {Kane}\ \emph {et~al.}(1997)\citenamefont {Kane},
  \citenamefont {Balents},\ and\ \citenamefont {Fisher}}]{Kane1997}%
  \BibitemOpen
  \bibfield  {author} {\bibinfo {author} {\bibfnamefont {C.}~\bibnamefont
  {Kane}}, \bibinfo {author} {\bibfnamefont {L.}~\bibnamefont {Balents}}, \
  and\ \bibinfo {author} {\bibfnamefont {M.~P.~A.}\ \bibnamefont {Fisher}},\
  }\href {\doibase 10.1103/PhysRevLett.79.5086} {\bibfield  {journal} {\bibinfo
   {journal} {Phys. Rev. Lett.}\ }\textbf {\bibinfo {volume} {79}},\ \bibinfo
  {pages} {5086} (\bibinfo {year} {1997})},\ \Eprint
  {http://arxiv.org/abs/9708054v1} {arXiv:9708054v1 [arXiv:cond-mat]}
  \BibitemShut {NoStop}%
\bibitem [{\citenamefont {Egger}\ and\ \citenamefont
  {Gogolin}(1997)}]{Egger1997}%
  \BibitemOpen
  \bibfield  {author} {\bibinfo {author} {\bibfnamefont {R.}~\bibnamefont
  {Egger}}\ and\ \bibinfo {author} {\bibfnamefont {A.~O.}\ \bibnamefont
  {Gogolin}},\ }\href {\doibase 10.1103/PhysRevLett.79.5082} {\bibfield
  {journal} {\bibinfo  {journal} {Phys. Rev. Lett.}\ }\textbf {\bibinfo
  {volume} {79}},\ \bibinfo {pages} {5082} (\bibinfo {year} {1997})},\ \Eprint
  {http://arxiv.org/abs/9708065} {arXiv:9708065 [cond-mat]} \BibitemShut
  {NoStop}%
\bibitem [{\citenamefont {Heedt}\ \emph {et~al.}(2016)\citenamefont {Heedt},
  \citenamefont {Prost}, \citenamefont {Schubert}, \citenamefont
  {Gr{\"{u}}tzmacher},\ and\ \citenamefont {Sch{\"{a}}pers}}]{Heedt2016}%
  \BibitemOpen
  \bibfield  {author} {\bibinfo {author} {\bibfnamefont {S.}~\bibnamefont
  {Heedt}}, \bibinfo {author} {\bibfnamefont {W.}~\bibnamefont {Prost}},
  \bibinfo {author} {\bibfnamefont {J.}~\bibnamefont {Schubert}}, \bibinfo
  {author} {\bibfnamefont {D.}~\bibnamefont {Gr{\"{u}}tzmacher}}, \ and\
  \bibinfo {author} {\bibfnamefont {T.}~\bibnamefont {Sch{\"{a}}pers}},\ }\href
  {\doibase 10.1021/acs.nanolett.6b00414} {\bibfield  {journal} {\bibinfo
  {journal} {Nano Lett.}\ }\textbf {\bibinfo {volume} {16}},\ \bibinfo {pages}
  {3116} (\bibinfo {year} {2016})}\BibitemShut {NoStop}%
\end{thebibliography}
\bibliographystyle{apsrev4-1}

\end{document}